\documentclass[aps,prb,twocolumn,superscriptaddress,eqsecnum,nofootinbib,showpacs]{revtex4}
\pdfoutput=1
\usepackage{amssymb}
\usepackage{graphicx}
\usepackage{epsf,epsfig}
\usepackage{bm}
\usepackage{bbm}
\usepackage{amsfonts}

\newcommand{\be}{\begin{eqnarray}}
\newcommand{\ee}{\end{eqnarray}}
\newcommand{\ba}{\begin{array}}
\newcommand{\ea}{\end{array}}
\newcommand{\no}{\nonumber}

\newcommand{\eps}{\varepsilon}

\newcommand{\bfr}{{\bf r}}

\newcommand{\bfB}{{\bf B}}

\newcommand{\bfA}{{\bf A}}

\newcommand{\erfc}{\mbox{erfc}}

\begin{document}

\title{Fluctuation persistent current in small superconducting rings}
\author{Georg Schwiete}
\email{schwiete@zedat.fu-berlin.de}
\affiliation{Dahlem Center for Complex Quantum Systems and Institut f\"ur Theoretische Physik,
Freie Universit\"at Berlin, 14195 Berlin, Germany}
\author{Yuval Oreg}
\affiliation{Department of Condensed Matter Physics, The Weizmann Institute of Science, 76100
Rehovot, Israel}
\date{\today}

\begin{abstract}
We extend previous theoretical studies of the contribution of
fluctuating Cooper pairs to the persistent current in
superconducting rings subjected to a magnetic field. For
sufficiently small rings, in which the coherence length $\xi$
exceeds the radius $R$, mean field theory predicts the emergence of a flux-tuned quantum critical point separating metallic and superconducting phases near half-integer flux through the ring. For larger rings with $R\gtrsim \xi$, the transition temperature is periodically reduced, but superconductivity prevails at very low temperatures. We calculate the fluctuation persistent current in different regions of the metallic phase for both types of rings. Particular attention is devoted to the interplay of the angular momentum modes of the fluctuating order parameter field. We discuss the possibility of using a combination of different pair-breaking mechanisms to simplify the observation of the flux-tuned transition in rings with $\xi>R$.

\end{abstract}
\pacs{74.78.Na, 73.23.Ra, 74.25.Ha}
\maketitle


\section{\label{sec:Introduction}Introduction}

The study of superconducting fluctuations has already a long history, for a comprehensive review see Ref.~\onlinecite{Larkin05}. When approaching the superconducting phase from the metallic side, for example by lowering the temperature $T$, precursors of superconductivity reveal themselves long before the superconducting state is fully established. In this regime, electrons form Cooper pairs only for a limited time. Being charged objects themselves, the Cooper pairs participate in charge transport. At the same time the density of states of the unpaired electrons is reduced. These simple qualitative arguments already indicate that superconducting fluctuations can affect both transport and thermodynamic properties of the metal \emph{outside} the superconducting phase. Detailed studies of these effects have been conducted in different contexts.\cite{Larkin05,Tinkham96} It is well known, for example, that fluctuation effects are more pronounced when the effective dimensionality of the superconductor is reduced or in the presence of disorder.

\begin{figure}
\centerline{\includegraphics[width=7.5cm]{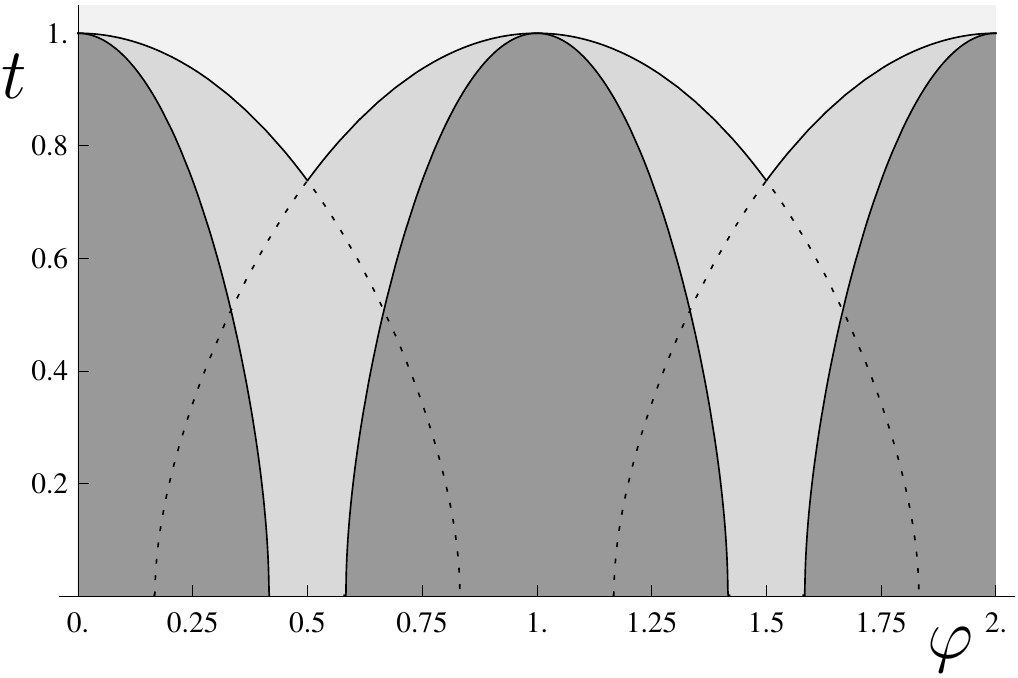}}
\caption{Mean field phase diagram. $T_{c\varphi}$ separates the metallic (high $T$) and the superconducting (low $T$) phase as a function of the flux $\varphi=\phi/\phi_0$
through the ring.
The transition line is determined by the condition $\mathcal{L}^{-1}_{00}=0$, cf. Eq.
(\ref{eq:fluctprop}). The superconducting phase for small rings with effective
radius $r=R/\xi < 0.6$ is shown in dark gray. Mean field theory predicts a full reduction of
$T_c$ for fluxes between $\varphi_{c0} \approx 0.83 r$ and
$1-\varphi_{c0}$ near $\varphi=1/2$. In this article we focus on fluctuations in the normal phase for these small rings. The superconducting phase for larger rings with $r\gtrsim 1$ is shown in light gray. $T_c$ is periodically reduced as a function of the flux $\varphi$, but superconductivity prevails at low temperatures.
The dotted lines gives $T_{n\varphi}$ defined below Eq. (\ref{eq:fullF}) for $n\in\{0,1,2\}$.
For $r\gtrsim 1$ it is well approximated by
the formula $T_{c\varphi}\approx T_{c0}(1-\varphi^2/r^2)$. The
phase diagram is periodic in $\varphi$ with period 1 for vanishing
ring thickness.  \label{fig:Tc}}
\end{figure}

In bulk superconductors the transition temperature $T_c$ can be partially or even completely suppressed by various pair-breaking
mechanisms, most notably by applying a magnetic field or introducing magnetic impurities. An additional pair-breaking mechanism can become effective in doubly connected superconductors like superconducting rings or cylinders, when they are threaded by a magnetic flux $\phi$. In this case one observes so-called Little-Parks
oscillations,\cite{Little62} the transition temperature $T_c$ is \emph{periodically} reduced as a function of $\phi$. Due to the periodicity, it is immediately evident that this effect is qualitatively different from the mere suppression of superconductivity by a magnetic field in bulk superconductors.
The period of the oscillations is equal to~$1$ as a function of the reduced flux $\varphi=\phi/\phi_0$, where the superconducting flux quantum is $\phi_0=\pi/e$,\cite{Units} see Fig.~\ref{fig:Tc}. The maximal $T_c$ reduction occurs when $\varphi$ takes half-integer values.

The magnitude of the $T_c$ reduction is size-dependent. It is convenient to measure the ring radius $R$ in units of the zero-temperature coherence length $\xi$ and to define $r=R/\xi$. A representative mean field phase diagram is displayed in Fig.~(\ref{fig:Tc}) for two rings of different size. As we see in Fig.~\ref{fig:Tc}, mean field theory predicts a moderate $T_c$ reduction for moderately small rings with $r\gtrsim 1$. Most strikingly, it also shows that for very small rings or cylinders with $r<0.6$ the transition temperature is expected to be equal to zero in a finite interval close to half-integer fluxes [This regime is sometimes called the \emph{destructive} regime.] Correspondingly, a flux-tuned quantum phase transition is expected to occur in these rings or cylinders at a critical flux $\varphi_{c0}$. The mean field transition line can be found from Eq.~(\ref{eq:tcvarphi}) to be discussed below.

As is well known, superconducting rings threaded by a magnetic flux $\phi$ support a dissipationless persistent current.\cite{Tinkham96,Imry02} In this article, we
study theoretically the \emph{fluctuation} persistent current in different regions of the phase diagram, both for rings with $r\gtrsim 1$ and moderate $T_c$ reduction and for rings with $r\lesssim 0.6$ with strong $T_c$ suppression. In particular, we will study in detail the large fluctuation persistent current $I$ which occurs even at fluxes for which $T_c$ is reduced to zero while the system has a finite resistance. A short account of the most important results of this study has already been presented in Ref.~\onlinecite{Schwiete09a}. Here we extend our study to different regions of the phase diagram, discuss the results in a broader context, and include details of the derivations.

This study is motivated by recent experiments that are significant to our understanding of fluctuation phenomena in
superconductors with doubly-connected geometry. Koshnick et al.~\cite{Koshnicklong07} measured the persistent current in small superconducting rings
with $r\gtrsim 1$, for the smallest rings under study $T_c$ was reduced by approximately $6\%$. Superconducting fluctuations in rings of this size are by now well understood, both experimentally and theoretically.\cite{Koshnicklong07,Oppen92,Daumens98,Schwiete09a} This is not so for smaller rings with $r<0.6$. Strong Little-Parks oscillations for cylinders with $r<0.6$, for which $T_c$ is reduced to zero near half-integer flux, have been observed in a transport measurement on superconducting cylinders.\cite{Liu01} It has so far, however, not been possible to measure the persistent current in superconducting \emph{rings} close to the flux-tuned quantum critical point.

The difficulty to access the destructive regime for rings is that the experiments require a high sensitivity of the measurement device as well as low temperatures. At the same time, the magnetic field should be strong enough to produce a sufficiently large flux penetrating the small ring area. We address this issue in this manuscript by discussing the possibility that a combination of different pair-breaking mechanisms can lead to progress in this direction. More specifically, we consider the combined effects of the magnetic flux through the ring's center on the one hand and of magnetic impurities and/or the magnetic field passing through the bulk material of the ring on the other hand. By direct calculation, we further explore how the presence of the quantum critical point influences the persistent current away from the quantum critical point, e.g., for temperatures of the order of $T_{c}^0\equiv T_c(\varphi=0)$.

The literature on superconductivity in systems with doubly connected geometry is extensive. We would like to point out a number of works, where related phenomena have been discussed. The possibility of finding complete suppression of superconductivity near half integer flux in small superconducting rings was pointed out by de~Gennes.\cite{DeGennes81a,DeGennes81} The phase diagram of superconducting cylinders was considered in Ref.~\onlinecite{Groff68} taking into account the interplay of pair-breaking effects caused by the flux on the one hand and the magnetic field penetrating the walls (of finite width) on the other hand. We will use their results when we discuss the influence of finite-width effects on the phase diagram.

A detailed study of the fluctuation persistent current in rings for which $T_c$ is reduced to zero by magnetic impurities (at any value of the flux) can be found in Refs.~\onlinecite{Bary08} and \onlinecite{Bary09}. These works address a long standing puzzle related to the observation of an unexpectedly large persistent current in copper rings.\cite{Levy90} It is suggested that these rings contain a finite amount of magnetic impurities, which suppress superconductivity and cause the rings to remain in the normal state even at low temperatures. Denoting the scattering rate on the magnetic impurities by $1/\tau_s$, there is a critical rate $1/\tau_{sc}$ and an associated quantum critical point that separates the superconducting from the normal phase. If the measurements are performed on rings with a scattering rate that is larger but close to $1/\tau_{sc}$, the corresponding fluctuations can lead to large currents in the rings. In contrast, in the parallel work of Ref.~\onlinecite{Schwiete09a} and in the present manuscript we consider the opposite case, in which the phase transition is primarily tuned by the magnetic flux, so that for vanishing flux and low temperatures the ring is in the superconducting state. We examine the influence of additional \emph{weak} pair-breaking effects on the phase diagram and how they can help to experimentally observe the flux-tuned quantum phase transition.

In the experiment of Ref.~\onlinecite{Liu01} on cylinders it was observed that near half-integer flux the resistance $R$ along the cylinder drops as $T$ decreases and then saturates for the lowest temperatures. In a later experiment \cite{Wang05}, regular step-like features where additionally observed in the $R-T$ diagram and interpreted as being due to a separation into normal and superconducting regions along the cylinder. A number of theoretical works addresses the issue of transport in small superconducting-cylinders. In Refs.~\onlinecite{Lopatin05} and \onlinecite{Shah07} the perturbative fluctuation contribution to the conductivity of long superconducting \emph{cylinders} near a flux-tuned quantum critical point was discussed as a particular example for \emph{transport} near a pair-breaking transition. In a broad sense, the general approach is similar to ours, but the considered system has a different dimensionality and the work discusses transport, while we study a thermodynamic property. It has been suggested in Ref.~\onlinecite{Lopatin05} that the observable regime in the experiment\cite{Liu01} is dominated by
thermal fluctuations and that at even lower $T$ an upturn of $R$ could be expected. To the best of our knowledge, so far no detailed comparison between theory and experiment is available. The observed saturation in the experiment\cite{Liu01} corresponds to a strong reduction of the normal resistance and as such lies outside the region of validity of the perturbative approach. The role of inhomogeneities along the cylinder axis has been further emphasized in Ref.~\onlinecite{Vafek05}. In Ref.~\onlinecite{Dao09} a mean field model was proposed that takes into account inhomogeneities caused by a variation of parameters like the mean free path or the width along the cylinder axis and a specific profile was found that would quantitatively fit the experimental phase diagram both as far as the saturation and the step-like features are concerned. Fluctuation effects, on the other hand, where neglected. It seems likely that a complete description would have to include both inhomogeneities and fluctuations, which is very demanding. In this respect, the situation with rings is more advantageous. Since the typical size of inhomogeneities is larger or of the order of the coherence length, they are unlikely to play a role for the superconducting rings under study here and we may focus on fluctuation effects only. We make detailed predictions for flux and temperature dependence of the resulting fluctuation persistent current. As mentioned before, due to experimental difficulties, (to the best of our knowledge) no measurements of the destructive regime are available for rings yet.

Fluctuation effects in moderately small superconducting rings ($r\gtrsim 1$) were studied in Refs.~\onlinecite{Daumens98} and \onlinecite{Buzdin02} with particular emphasis on the regime of strong fluctuations near the transition. These works introduce the idea that the strong fluctuation regime near the thermal transition can be described in terms of one or two coupled angular momentum modes of the order parameter field in the classical GL functional. We will make use of this idea and derive more detailed results for the persistent current and susceptibility close to integer and half-integer fluxes.

Ref.~\onlinecite{Oppen92} studies superconducting fluctuations in rings near the thermal transition using a numerical approach based on the mapping of the classical GL-functional onto the problem of solving an effective Schr\"odinger equation\cite{Scalapino72}. The results of this approach agree well with the experiment of Koshnick et al.\cite{Koshnick07} This approach works nicely in cases where many angular momentum modes of the order parameter field give a sizeable contribution to the persistent current, but can become cumbersome in the opposite limit (see Ref.~\onlinecite{Koshnick07}). In this sense, it is complementary to the approach used in this manuscript (as well as in Refs.~\onlinecite{Daumens98, Buzdin02, Schwiete09a}), which is well suited for rings that are so small that only one or two angular momentum modes are important.

The role of the back-action effects caused by the self-induction of cylinders and rings was discussed in this context in Refs.~\onlinecite{Fink80} and \onlinecite{Fink86}. While we will discuss finite thickness effects, we will generally assume that the self-induction and the persistent currents of the \emph{rings} under study in this article are sufficiently small so that back-action effects can be neglected.

The article is organized as follows. Section \ref{sec:thermal} is devoted to the persistent current in rings with $r\gtrsim 1$. Some calculational details are presented in appendix \ref{app:twozero}. In section \ref{sec:quantum} we study in detail the fluctuation persistent current in different regions of the phase diagram for rings with $r\lesssim 0.6$, including the vicinity of the quantum critical point. Details of the calculation are relegated to appendix \ref{app:cplane}. In section \ref{sec:discussion} we discuss different ways to reduce $T_c$ to zero in rings with finite width or by introducing magnetic impurities.

\section{Thermal transition for rings with $r\gtrsim 1$}
\label{sec:thermal}
In this section we will discuss the description of rings with only a moderate suppression of $T_c$, i.e., rings for which $r\gtrsim 1$. A condensed discussion has already been presented in Ref.~\onlinecite{Schwiete09a}. Here we take the opportunity to provide additional information. For the rings with $r\gtrsim 1$ the superconducting transition occurs at a finite temperature and fluctuations can be described with the help of the classical GL functional in which the order parameter field is static. In the imaginary time formalism this amounts to neglecting order parameter field components with finite Matsubara frequency in the functional\cite{Larkin05} (a formal justification will be given in the first paragraph of Sec.~\ref{subsec:Fluctuation} below). This simplified description is valid close to the transition line in the $T_c-\varphi$ phase diagram.

\subsection{Ginzburg-Landau functional}

The starting point for our discussion of superconducting fluctuations in rings with $r\gtrsim 1$ is the classical GL functional\cite{Ginzburg50}
\be
\label{eq:Fclassical}
\mathcal{Z}&=&\int D(\psi,\psi^*) \exp\left(-\mathcal{F}/T\right)\\
\mathcal{F}&=&\mathcal{F}_N+\int d\bfr
\left(a|\psi(\bfr)|^2\right.\no\\
&&\left.+\frac{b}{2}|\psi(\bfr)|^4+\frac{1}{4m}\big|(-i\nabla-2e{\bf
A}(\bfr))\psi(\bfr)\big|^2\right)\qquad\no
\label{eq:classicalF}
\ee
$\mathcal{F}_N$ describes the the normal (non-superconducting) part of the free energy. It gives rise to a normal component of the persistent current.\cite{Cheung89} Since we are mainly interested in a regime close to the transition, however, the fluctuation contribution is much larger and we will not discuss $\mathcal{F}_N$ further.
At the mean field level, the sign change of the quadratic form in $\psi$ signals the onset of the superconducting phase, which motivates the parametrization
$a=\alpha T^0_c\eps$, where $\eps=\frac{T-T^0_c}{T^0_c}$. A characteristic length scale, the (zero temperature) coherence length, can be identified $\xi=1/\sqrt{4m\alpha T^0_c}$. The microscopic theory for disordered superconductors\cite{Gorkov59} gives rise to the following relations
\be
{\alpha^2}/{b}={8\pi^2\nu}/{7\zeta(3)},\quad \xi^2={\pi D}/{8T_c}
\ee
where $D$ is the diffusion coefficient. The normalization of $\psi$ allows for a certain arbitrariness, this is why only the ratio of $\alpha^2$ and $b$ is fixed.

The quartic part of the functional stabilizes the system once it is tuned below the transition temperature. Above this temperature, the quartic term gives only a small contribution to thermal averages, except in the very vicinity of the transition, the so-called Ginzburg region. As long as one stays outside of this region on the metallic side, one can restrict oneself to a quadratic (Gaussian) theory (i.e. neglect the quartic term), which is much easier to handle theoretically, but becomes unreliably close to the transition where the quadratic theory becomes unstable. Importantly, even above the transition temperature, in the normal region of the mean field phase diagram, the average of $|\psi|^2$ with respect the functional $\mathcal{F}$ is finite.

After this preparation, we turn to the description of superconducting rings.
When the superconducting coherence length $\xi(T)=\xi/\sqrt{\eps}$ and the
magnetic penetration depth $\lambda(T)$ are much larger than the
ring thickness, the system is well described by a one-dimensional
order parameter field $\psi$, albeit with periodic boundary conditions \cite{Imry02}.
In order to account for these boundary conditions, it is convenient to introduce angular momentum modes as
\be
\psi(\vartheta)=\frac{1}{\sqrt{V}}\sum_n\psi_n\;\mbox{e}^{in\vartheta}.
\ee
$V=2\pi R S_\perp$ is the volume of the ring, $S_\perp$ the cross-section of the wire forming the ring. The vector potential can be chosen as $\bfA=\bfB\times\bfr/2$, the integration as $\int d\bfr\rightarrow S_\perp R \int d\phi$. Then, the free energy functional
takes the form
\be
\mathcal{F}=\sum_n a_{n\varphi}|\psi_n|^2+\frac{b}{2V}\sum_{nmkl}\delta_{n+k,l+m}\psi_n\psi^*_m\psi_k\psi_l^*
\label{eq:fullF}
\ee
where $a_{n\varphi}=a+(n-\varphi)^2/2mR^2$. Let us make three important observations. First, the free energy functional is flux dependent. As a consequence a persistent current can flow in the ring. Second, the functional is periodic in the reduced flux $\varphi$ with period 1. Correspondingly, the same is true for all thermodynamic quantities derived from the the GL functional. This property holds strictly speaking only in the idealized limit of a one-dimensional ring. In reality, the external magnetic field also penetrates the superconductor and provides an additional mechanism for the suppression of superconductivity. We will come back later to this point in section~\ref{sec:discussion}. The third observation is that now the kinetic energy of the Cooper pairs vanishes only when the reduced flux $\varphi$ takes integer values. Otherwise it gives a finite contribution to the quadratic part of the functional and therefore the transition takes place at a temperature $T_c(\varphi)$ that is in general reduced with respect to $T_c^0$ of the bulk material. Let us parameterize $a_{n\varphi}=\alpha T_c^0\eps_{n\varphi}$. Then $\eps_{n\varphi}=(T-T_{n\varphi})/{T^0_c}$ and $a_n$ change sign at a temperature
\be
T_{n\varphi}=T_c^0[1-(n-\varphi)^2/r^2].
\ee
This temperature can loosely be interpreted as the transition temperature of the $n$th angular momentum mode $\psi_n$. The mean field transition for the ring occurs at
$T_{c\varphi}$ that is equal to the maximal $T_{n}$ for given $\varphi$, i.e., at the point where the first mode becomes superconducting when lowering the temperature (cf. Fig. \ref{fig:Tc}).

In the subsequent discussion, the parameter $\Lambda=1/r^2Gi$ will play a crucial role. Its relevance is now easily understood. $1/r^2$ is a measure for the typical spacing between the transition temperatures $T_n$ for different modes, since e.g. $[T_0(\varphi)-T_1(\varphi)]/T_c^0=(1-2\varphi)/r^2$ (compare Fig. \ref{fig:Tc}). This spacing can be compared to the typical width of the non-Gaussian fluctuation region, $Gi$. Only in this region fluctuations are strong. The zero-dimensional Ginzburg parameter $Gi$ of relevance here is given as\cite{Larkin05}
\be
Gi=\sqrt{\frac{2b}{\alpha^2T_c^0 V}}=\sqrt{\frac{7\zeta(3)}{4\pi^2\nu T_c^0 V}},
\ee
where $\nu$ is the density of states at the Fermi level and $V$ is the volume of the ring. The parameter $\Lambda=1/r^2Gi$ determines whether a description in terms of a few modes only is a good approximation in the critical regime (for $\Lambda \gg 1$) or not. Defining the dimensionless conductance of the ring as $g=R_Q/R_\circ=2e^2\nu D R_Q V/(2\pi R)^2$, $R_Q=\pi /e^2$, one can find an alternative expression, $\Lambda\approx 5\sqrt{g}/r$. We state this alternative (but equivalent) expression for $\Lambda$, because it might be more convenient for estimates when performing an experiment.

An analytic computation of the functional integral necessary to obtain $\mathcal{Z}$ is in general not possible and one has to resort to approximation schemes.
If the spacing is large ($\Lambda \gg 1$) and one is interested in the region of strong fluctuations close to the transition temperature $T_c(\varphi)$, an effective theory including only one angular momentum mode $\psi_n$ for $n\sim \varphi$ is applicable,
\be
\label{eq:Fn}
\mathcal{F}_n\sim a_n|\psi_n|^2+\frac{b}{2V}|\psi_n|^4.
\ee
This is so, since in this case the temperature $T\sim T_c(\varphi)$ lies far above the individual transition temperatures $T_m(\varphi)$ ($m\ne n$) of all other modes $\psi_m$ and they will give only a small contribution when calculating observables. This is very convenient, because in this case one comes to the zero-dimensional limit of the GL functional.\cite{Muhlschlegel72,Larkin05} The partition function based on this free energy functional can be calculated exactly and all thermodynamic quantities derived from it.

Clearly, for half-integer values of the flux the spacing between two adjacent modes always goes to zero and due to this degeneracy at least two modes are required for the description. Let us choose for definiteness the example $0<\varphi<1$, then one may work with
\be
\label{eq:2}
\mathcal{F}_{01}=\sum_{i=0,1}a_i|\psi_i|^2+\frac{b}{2V}[|\psi_0|^4+|\psi_1|^4+4|\psi_0|^2|\psi_1|^2]\quad.
\ee
In this situation the parameter $\sqrt{g}/r\approx 1/5r^2Gi$ is still useful, because if it is large, additional modes need not be taken into account and a two-mode description is valid.

If $\sqrt{g}/r$ is not exceedingly large, the small contribution of the remaining modes can easily be accounted for by using the Gaussian approximation for them. One should, however, not forget that the presence of the dominant mode(s) can influence the effective transition temperatures of all others via the quartic term. As an example, let us write the resulting effective action for the modes with $n\ne 0$ assuming that $\varphi\sim 0$ and the mode with $n=0$ is the dominant one,
\be
\label{eq:Feff}
\mathcal{F}^{e\!f\!f}_{n\ne 0}=\sum_{n\ne 0} (a_n+\alpha^2Gi^2T_c^0|\psi_0|^2)|\psi_n|^2.
\ee
Essentially the same argument was first presented in Ref.~\onlinecite{Daumens98}. If temperatures are sufficiently high $T\gg T^0_c(1+Gi)$ the quartic term may be dropped altogether and one may work with a purely Gaussian theory $\mathcal{F}_n\approx \sum_n a_{n\varphi}|\psi_n|^2$.

\subsection{Persistent current and susceptibility}

The persistent current $I$ is found from the free energy $F=-T\ln \mathcal{Z}$ by differentiation $I=-\partial F/\partial \phi$. The normalized current is given by
\be
\label{eq:Idef} i=I/(T_c^0/\phi_0)=\sum_{n=-\infty}^{\infty}
\frac{2\alpha}{ r^2}\left(n-\varphi\right)\left\langle
|\psi_n|^2\right\rangle.
\ee
The averaging is performed with respect to the functional $\mathcal{F}$ in Eq.~(\ref{eq:fullF}). Just as the free energy functional $\mathcal{F}$, the persistent current $i$ is periodic in the flux $\varphi$ with period one. Since it is also an odd function of $\varphi$, the persistent current vanishes when $\varphi$ takes integer or half-integer values.

\paragraph{Case $\varphi\approx n$, $T\approx T_c$:}
As pointed out above, the most important contribution in the regime of non-Gaussian fluctuations close to integer fluxes comes from the angular momentum mode $\psi_n$ with the highest transition temperature $T_{n\varphi}$. One may then approximate Eq.~(\ref{eq:fullF}) by a single-mode and calculate with $\mathcal{F}_n=a_n|\psi_n|^2+\frac{b}{2V}|\psi_n|^4$. This is the 0d limit of the GL functional~\cite{Muhlschlegel72} already introduced above. In this limit, Eq.~(\ref{eq:Idef}) gives
\be
i_n=4\Lambda(n-\varphi)f(x_n)\label{eq:scaling}\;\; \mbox{ for } \varphi \approx n.
\ee
Here $x_n=\eps_n/Gi$ and the function
\be
f(x)=\frac{\exp(-x^2)}{\sqrt{\pi}\mbox{erfc}(x)}-x
\ee
is defined with the help of the conjugated error function.\cite{Abramowitz72} All persistent current measurements will fall on the same curve, if the persistent current -- measured in suitable units $i =I/(T_c^0/\phi_0)$ -- and the reduced temperature $\varepsilon_\varphi=(T-T_{c\varphi})/T^0_c$ are scaled as
\be
\label{eq:iscaling}
i \rightarrow i \frac{r}{\sqrt{g}}, \quad \epsilon_\varphi \rightarrow \epsilon_\varphi \ r \sqrt{g}.
\ee
This relation can serve as a valuable guide in characterizing different rings in experiments.

\paragraph{Gaussian theory for $T\gg T_c^0$, estimate for $T\ll T_c^0$:}
It is possible to make contact with the Gaussian and the mean field results using the asymptotic expansion of the conjugated error function
\be
\sqrt{\pi}x\mbox{erfc}(x)\approx \exp(-x^2)(1-1/(2x^2))\quad (x\rightarrow \infty)
\ee
and the limit $\mbox{erfc}(x)\rightarrow 2$ for $x\rightarrow -\infty$. Far above $T_c$ one obtains as a limiting case the Gaussian result for a single mode $i_n\approx 2(n-\varphi)/r^2\eps_{n\varphi}$,
that can also be obtained directly by neglecting the quartic term in the GL functional. In this form, however, it is of limited use, since for the temperatures in question one should sum the contribution of all modes. Indeed, in this case one can use the relation $\left\langle|\psi_n|^2\right\rangle\sim T_c^0/ a_n$ ($a_n$ was defined below Eq.~\ref{eq:fullF}) and perform the sum in Eq.~(\ref{eq:Idef}) to obtain a result that is valid at arbitrary fluxes~\cite{Ambegaokar90}
\be
i(\varphi)=\frac{-2\pi\sin(2\pi\varphi)}{\cosh(2\pi\sqrt{\eps}r)-\cos(2\pi\varphi)}
\label{eq:Ambgauss}
\ee
One of the main features of this result besides the periodicity in $\varphi$ is the exponential decay of the persistent current as a function of temperature for  $\eps>1/(2\pi)^2r^2$, which is due to a mutual cancelation of the contributions of many modes to the persistent current. Turning back to Eq.~(\ref{eq:scaling}), we see that far below $T_c$ one recovers the mean field result
\be
i_{MF} \equiv \frac{-4}{r^2Gi^2}\eps_{n\varphi}(n-\varphi),
\ee
which gives an estimate for the persistent current in the superconducting regime. In this approximation the current grows linearly with $|T-T_c|$ and as soon as $|\eps|\gg Gi+(n-\varphi)^2/r^2$ (i.e. $|x_n|\gg 1$) one expects a sawtooth-like behavior as a function of the flux (i.e. linear dependence from $\varphi=n-1/2$ to $\varphi=n+1/2$ passing through zero at integer $n$) with a discontinuous jump at half integer $\varphi$. Both the Gaussian and the mean field result are reliable only outside the region of strong fluctuations, the persistent current $i_n$ in Eq.~(\ref{eq:scaling}) covers this region and interpolates smoothly between them.

\paragraph{Case $\varphi \approx n+1/2$, $T\approx T_c$:}
At half integer values of $\varphi$, the transition temperatures for two modes become equal. In the vicinity of this point in the phase diagram the two dominant modes influence each other, their \emph{coupling} becomes crucial. We discuss the case $\varphi\approx 1/2$ for definiteness, and use the form of the free energy functional already displayed in Eq.~(\ref{eq:2}).

Calculation of the persistent current in the presence of the coupling requires a generalization of the approach used for the single mode case.\cite{Daumens98} Explicit formulas for the persistent current are derived and displayed in appendix \ref{app:twozero} for the sake of completeness. For a graphical illustration see Fig.~2 of Ref.~\onlinecite{Schwiete09a}. We define  the susceptibility as $\chi=-\partial I/\partial\phi$. Differentiating the expression for $i_2$ (see appendix~\ref{app:twozero}) one obtains
\be
\overline{\chi}_{\varphi=1/2}&=&\frac{\chi}{T^0_c/\phi_0^2}=4\Lambda\;g_1\left(x\right)-4\Lambda^2\;g_2\left(x\right),
\label{eq:f2}
\ee
where \cite{Scales}
\be
x\equiv x_0(1/2)=x_1(1/2)=\frac{\eps}{Gi}+\frac{1}{4}\Lambda.
\ee Note that $x=0$ at the mean field transition for $\varphi=1/2$, i.e. the condition $x=0$ defines $T_{c,1/2}=T^0_c(1-1/4r^2)$. The dimensionless smooth functions $g_n$ are defined by the relations
\be
g_1(x)&=&\frac{1}{2J(x)}\textrm{e}^{\frac{1}{3}x^2}\textrm{erfc}(x)-\frac{2x}{3}\\
g_2(x)&=&\frac{3}{2\sqrt{\pi}J(x)}\;\textrm{e}^{-\frac{2}{3}x^2}-\frac{3x}{2J(x)}\textrm{e}^{\frac{1}{3}x^2}\;\textrm{erfc}(x)-1,\no\\
\ee
where $J(x)=\int_x^\infty dt\;\textrm{e}^{\frac{1}{3}t^2}\;\textrm{erfc}(t)$. Exact results can be given for the functions $g_n$ at the transition, i.e. for $x=0$. These give a useful estimate for the magnitude inside the fluctuation region, $g_1(0)=\sqrt{{\pi}}/[{2\sqrt{3}\mbox{arctanh}\left({1}/{\sqrt{3}}\right)}]\approx 0.78$ and
$g_2(0)={\sqrt{3}}/[2\mbox{arctanh}\left({1}/{\sqrt{3}}\right)]-1\approx 0.315$. For large $\Lambda =1/r^2Gi\approx 5\sqrt{g}/r$ one can neglect the first term in Eq.~(\ref{eq:f2}). Then one obtains $\overline{\chi}_{1/2}=-\;4\Lambda^2\; g_2\left(x\right)$. For the susceptibility close to integer flux one easily obtains $\overline{\chi}_{0}=4\Lambda f(x_0)$  from Eq.~(\ref{eq:scaling}). Comparing to the expression for $\overline{\chi}_{1/2}$, we find
\be
\chi_{1/2}/\chi_0 \approx -2.7 \sqrt{g}/r. \label{eq:chi}
\ee
Experimentally, a strong enhancement of the magnetic susceptibility near $\varphi=1/2$ compared to $\varphi \approx
0$ was observed\cite{Koshnick07} and Eq.~(\ref{eq:chi}) demonstrates that it is controlled by the parameter $\sqrt{g}/r$. If it is large, the current rapidly changes sign as a function of the flux at half-integer flux, leading to a saw-tooth like shape of $i_{\varphi}$. The full $T$ dependence of $\chi_{\varphi=1/2}$
is given in Eq.~(\ref{eq:f2}). The shape of the persistent current as a function of the flux has been discussed in more detail in Ref.~\onlinecite{Schwiete09a}. Let us just stress the main physical mechanism at work here. To be specific, we discuss the vicinity of $\varphi=1/2$. Close to $1/2$ there is a competition of two angular momentum modes, $\psi_0$ and $\psi_1$, that are almost degenerate. If one tunes to a slightly smaller flux, say, then the mode $\psi_0$ is dominant, because $a_0$ is smaller than $a_{1}$. The effect of the coupling term in this case is to further weaken the mode $\psi_1$ (this can be seen from an effective action in the form displayed in Eq.~\ref{eq:Feff}.), for a
repulsive interaction the dominant mode suppresses the subdominant mode. Since the contribution to the persistent current of these two modes is opposite in sign, the result is an almost saw-tooth like shape of the persistent current.\cite{Schwiete09a}. Similar effects related to the competition of two order parameter fields have been discussed in the past, see, e.g., Ref.~\onlinecite{Imry74}.

\paragraph{Summary:}
To summarize, in this section we argued that for $\Lambda = 5 \sqrt{g}/r \gg 1$ one can use simple approximation schemes to calculate the persistent current and susceptibility near $T_c$. The significance of the parameter $\Lambda$ is as follows. If $\Lambda$ is large, the statistical weight of one angular momentum mode in the strong fluctuation region by far exceeds the weight of all other modes, unless two modes become degenerate. The degeneracy points are $\phi=n+1/2$. Our calculations were based on formula \ref{eq:Idef} which is a general expression for the fluctuation persistent current expressed in terms of the thermal averages $\left\langle|\psi_n|^2\right\rangle$. Near integer flux ($\varphi \approx n$) and for $T\approx T_c$ it is sufficient to include only one mode, which leads to formula \ref{eq:scaling}. The periodicity of the phase diagram is not crucial here. Instead, the relevant free energy (Eq.~\ref{eq:Fn}) has the same form as that of a small superconducting grain in the zero-dimensional limit.\cite{Muhlschlegel72} This is a drastic simplification compared to the original problem and implies a high degree of universality. With the appropriate scaling given in Eq.~\ref{eq:iscaling}, $i(\varphi)$-data measured for rings with various parameters should fall on the same curve. As the temperature increases or the flux comes closer to the degeneracy points, the restriction to only one mode is no longer a good approximation. Formula \ref{eq:Ambgauss}, which is valid for high temperatures, includes the Gaussian contribution of formula \ref{eq:scaling} \emph{and} all the other modes. It is therefore applicable for arbitrary fluxes. The shape of $i(\varphi)$ becomes more and more sinusodial as the temperature increases, this is the result of the combined contribution of many modes. Returning to the vicinity of $T_c(\varphi)$, in formula \ref{eq:f2} we calculate the contribution of two modes to the magnetic susceptibility at half integer fluxes ($\varphi\approx n+1/2$). Formula \ref{eq:chi} compares the magnetic susceptibility at half-integer and integer fluxes and reflects the large enhancement of the susceptibility near $\varphi\approx n+1/2$, which is controlled by \emph{two} fluctuating angular momentum modes. We interpret this enhancement as being due to the suppression of the subdominant by the dominant mode as a result of their interaction.~\onlinecite{Imry74} This interaction is induced by the quartic term of the GL functional. In table \ref{table:formulas} we summarize the expressions and analytical approximations for the fluctuation persistent current near the thermal transition together with their range of validity.

\setlength{\tabcolsep}{0.2cm}
\begin{table*}
\begin{tabular}{|c|c|c|}
\hline
Formula&Meaning&Range of validity\\
\hline
(2.10)&\begin{tabular}{cc}PC, general formula,
all $\psi_n$ contribute \end{tabular}&thermal transition\\
\hline
(2.11)&PC, strong fluctuations, one $\psi_n$ dominates &$\varphi\approx n$, $|T-T_c|\lesssim T_cGi$, $\Lambda\gg 1$
\\
\hline
(2.15)&PC, gaussian fluctuations, all $\psi_n$ contribute&$\varphi$ and $\Lambda$ arbitrary, $T\gg(1+Gi)T_c$\\
\hline
(2.16)&PC, mean field estimate, one $\psi_n$ dominates & $\varphi\approx n$, $T\ll T_c$, $\Lambda\gg 1$\\
\hline
(A5)&PC, strong fluctuations, two $\psi_n$ contribute& $\varphi\approx n+1/2$, $|T-T_c|\lesssim T_cGi$, $\Lambda \gg 1$\\
\hline
(2.17)&Susceptibility, strong fluctuations, two $\psi_n$ contribute& $\varphi\approx n+1/2$, $|T-T_c|\lesssim T_cGi$, $\Lambda \gg 1$\\
\hline
\end{tabular}
\caption{
An overview of different formulas for the \emph{fluctuation} persistent current (PC) near the thermal transition for rings with $r\gtrsim 1$. The range of validity for the analytical approximations is also estimated. $T_c$ is the (flux dependent) critical temperature, $\varphi = \phi/(h/2e)$ is the dimensionless flux through the interior of the ring and $Gi$ is the 0d-Ginzburg parameter defined in Eq.~(2.6). The dimensionless parameter $\Lambda$ has been introduced after  Eq.~(2.6) and can be much larger than $1$ for small rings. \label{table:formulas}}
\end{table*}

\section{Gaussian fluctuations for rings with $r<0.6$}
\label{sec:quantum} So far we have discussed the limit of
moderately small rings with $\protect{r=R/\xi>1}$. For these rings the
transition temperature is only weakly suppressed at finite flux
and the phase transition occurs at a finite temperature. We will
now discuss even smaller rings with $r<1$. For such rings the
theoretical description based on the classical GL functional we used
so far is not valid in large regions of the phase diagram, as it
is applicable only in a relatively small temperature interval
close to $T_c^0$. For rings with $r<1$, however, the transition
near half-integer flux can occur at temperatures far below
$T_c^0$, or even at vanishing temperatures. The theoretical
description for these rings can be developed in close analogy to
the general theory of pair-breaking transitions. In this section
we will first determine the mean field transition line and then
calculate the contribution of Gaussian fluctuations to the
persistent current outside the superconducting regime.

\subsection{Mean field transition line}
The partition function is conveniently formulated in terms of an
integral over a complex order parameter field $\Delta$ as
$\mathcal{Z}=\int D(\Delta,\Delta^*)\exp(-S)$, where \be
\mathcal{S}&=&T\sum_\omega \sum_n
\Delta^*(n,\omega)\mathcal{L}^{-1}(n,\omega)\Delta(n,\omega).
\label{eq:Ssimple}
\ee
The field $\Delta$ -- unlike the field $\psi$ in the classical GL
functional of Eq.~\ref{eq:Fclassical} -- is dynamical, $\omega_m=2\pi m T$ is a bosonic
Matsubara frequency. For a derivation of Eq.~\ref{eq:Ssimple} see, e.g., Chapter 6 of Ref.~\onlinecite{Larkin05}.  Let us note here that the field $\psi_0$ is
proportional to the static component $\Delta(\omega_n=0)$. A
neglect of fields with nonzero Matsubara frequencies can be
justified for the thermal transition, where it leads to the
classical GL functional, but is not justified near the quantum
phase transition. In full analogy to previous considerations,
$\Delta$ has been expanded in terms of angular momentum modes. The
fluctuation propagator $\mathcal{L}$ fulfills
\be
&&(\nu\mathcal{L})^{-1}(n,\omega)\label{eq:fluctprop}\\
&&=\ln\left(\frac{T}{T^0_c}\right)+\psi\left(\frac{1}{2}+\frac{\alpha_n+|\omega|/2}{2\pi
T}\right)-\psi\left(\frac{1}{2}\right). \no
\ee
Here we introduced the pair-breaking parameter
\be
\alpha_n=\eps_T(n-\varphi)^2/2
\ee
for the problem under consideration and $\nu$ is the density of states at the Fermi
energy. We use the notation $\eps_T=D/R^2$ and $\psi$ is the Digamma function.\cite{Abramowitz72} Since we are interested in the mean field transition line and in the Gaussian fluctuations on the normal side of the transition, the quartic and higher order terms in the action of Eq.~\ref{eq:Ssimple} may be safely neglected (see, e.g., the book \cite{Larkin05}).

As usual, the mean field transition occurs when
$\mathcal{L}^{-1}(n,\omega)$ changes sign first for arbitrary $n$
and $\omega$. Assuming that $|\varphi|<0.5$, this happens for
$n=0$, $\omega=0$, so that the condition for the mean field
transition reads $\mathcal{L}^{-1}(0,0)=0$. It defines an implicit
equation for $T_c(\varphi)$ \be
\ln\left(\frac{T_c(\varphi)}{T^0_c}\right)=\psi\left(\frac{1}{2}\right)-\psi\left(\frac{1}{2}+\frac{\alpha_0(\varphi)}{2\pi
T_c(\varphi)}\right) \label{eq:tcvarphi} \ee The larger
$\alpha_0$, the lower values of $T_c(\varphi)$ are required to
fulfill this relation. Eventually, for
\be
\alpha_0(\varphi_c)=2\pi
T_c^0\exp(\psi(1/2))={\pi T_{c}^0}/{2\gamma_E}
\ee
 the transition
temperature $T_c(\varphi)$ vanishes, i.e. one reaches a
(flux-tuned) quantum critical point. This happens for the critical
flux\cite{Units}
\be
\varphi_{c}=\pi r/(2\sqrt{2\gamma_E}),\quad \gamma_E\approx
1.78.
\ee
Due to the flux-periodicity of the phase
diagram, a quantum transition can only be observed in the ring
geometry if $\varphi_{c}<1/2$, which implies
$r<\sqrt{2\gamma_E}/\pi\approx 0.6$ (compare Fig.~\ref{fig:Tc}). Notice that this critical
value of $r=R/\xi$ is less restrictive than a naive
application of the quadratic approximation valid for $r \gg 1$
would suggest. The latter would give $1-(1/2r)^2=0 \Rightarrow r=1/2$.

\subsection{Persistent Current}
The formula for the persistent current
$I=\frac{T}{\phi_0}\partial_\varphi\ln\mathcal{Z}$ in the Gaussian
approximation reads \be
I=-\frac{T}{\phi_0}\sum_{n,\omega}\frac{\partial\mathcal{L}^{-1}(n,\omega)}{\partial\varphi}\mathcal{L}(n,\omega),
\label{eq:igaussgeneral} \ee
where we used Eq.~\ref{eq:Ssimple}. In order to gain a good qualitative
and quantitative understanding of the flux and temperature
dependence, the magnitude and relevance of the scales involved in
the problem, we will in the following derive simpler expressions
for the persistent current and discuss various limiting cases. A
particular emphasis will be put on the analysis of the persistent
current in the vicinity of the quantum critical point. This
approach will be corroborated by a direct numerical evaluation of
(\ref{eq:igaussgeneral}). When doing so, some care needs to be
exercised in order to correctly deal with the slow convergence
properties for large values of $|n|$ and $|\omega|$. As before, we
find it convenient to consider the dimensionless quantity
$i=I/(T_c^0/\varphi_0)$ and for definiteness analyze fluxes
$\varphi$ in the interval $(0,0.5)$. Since the persistent current
is periodic $i(\varphi+1)=i(\varphi)$ and odd
$i(\varphi)=-i(-\varphi)$ in the flux, this is sufficient to infer
the persistent current for arbitrary fluxes.

As is shown in appendix~\ref{app:cplane}, the persistent current
can be written as the sum of two contributions, $i=i_s+i_{ns}$.
The rationale behind this decomposition is the following: The
singular part $i_s$ diverges on the transition line
$\varphi_c(T)$. Outside the Ginzburg region on the normal side of
the transition, $i_s$ is still strongly flux and temperature
dependent. It describes by far the dominant contribution to the
fluctuation persistent current in the entire normal part of the
phase diagram. The nonsingular part $i_{ns}$, on the other hand,
displays a smooth flux dependence and is much smaller. $i_s$ and
$i_{ns}$ have opposite signs. We will mostly discuss $i_{s}$, for
explicit formulas for $i_{ns}$ we refer to appendix~\ref{app:cplane}. The expression for $i_s$ reads
\be i_s&=&-2\pi
t\sum_\omega \frac{\sin(2\pi\varphi)}{\cosh(2\pi z_s)-\cos(2\pi
\varphi)} \label{eq:ismain} \ee
Here, $z_s$ is defined as the
unique positive solution of the equation \be
\psi\left(\frac{1}{2}\right)-\ln\left(\frac{T}{T_c^0}\right)=\psi\left(\frac{1}{2}+\frac{|\omega|-\eps_Tz^2}{4\pi
T}\right), \ee for which the argument of the digamma function on
the right hand side is positive.
The formula for $i_s$ in Eq.~\ref{eq:ismain} can be viewed as a generalization of Eq.~\ref{eq:Ambgauss} to the entire normal part of the phase diagram (outside the Ginzburg regime).

In order to find a more convenient expression for $z_s$ it is
useful to define the function $\alpha_c(T)$ describing the phase
boundary in the $\alpha_0$--T phase diagram, i.e. $
\alpha_c(T)=\alpha_0(\varphi_c(T))$,
\be
\ln\left(\frac{T}{T^0_c}\right)=\psi\left(\frac{1}{2}\right)-\psi\left(\frac{1}{2}+\frac{\alpha_c(T)}{2\pi
T}\right). \label{eq:alphat} \ee One immediately reads off \be
z_s=\sqrt{(|\omega|-2\alpha_c(T))/\eps_T}, \ee
 Here, $\alpha_c(T)$
in Eq.~(\ref{eq:alphat}) is allowed to become negative as soon as
$T>T_c^0$. Note that $z_s$ is always real for $T>T_c^0$. This is
no longer true for $T<T_c^0$. In this regime it is instructive to
use the fact that $\alpha_c(T)$ and the temperature dependent
critical flux $\varphi_c(T)$ are closely related,
$\eps_T\varphi^2_c(T)=2\alpha_c(T)$. As a result,
\be
z_s=\varphi_c(T)\sqrt{|\omega|/2\alpha_c(T)-1},\quad (T<T_c^0).
\ee
Now $z_s$ is real for $\omega>2\alpha_c(T)$, but becomes purely
imaginary for $\omega<2\alpha_c(T)$. The fact that $z_s$ becomes
imaginary for small $|\omega|$ and $T<T_c^0$, but not for larger
$T>T_c^0$, is intimately related to the occurrence of the phase
transition. Indeed, the denominator in the expression \ref{eq:ismain} for $i_s$
vanishes for $\varphi=\varphi_c(T)$ at $\omega=0$, signaling the
onset of the superconducting regime. For reference, recall that
$\alpha_{c0}\equiv\alpha_c(0)={\pi T_{c}^0}/{2\gamma_E}\approx
0.88 T_c^0$.

\subsection{Fluctuation persistent current for $T>T_c^0$}
\label{subsec:Fluctuation}

Let us first make contact with the Gaussian result for larger
rings $r\gtrsim 1$ stated before in Eq.~(\ref{eq:Ambgauss}). For
$T\sim T_c^0$, when the logarithm on the left hand side of
Eq.~(\ref{eq:alphat}) is small, one can expand the Digamma
function on the right hand side. When keeping only the most
dominant term in the sum, the term with vanishing Matsubara
frequency, one easily finds $z_s\sim\sqrt{\eps}r$ and in this way
reproduces the result of Eq.~(\ref{eq:Ambgauss}) after setting
$t\approx 1$. The restriction to $\omega=0$ is justified in this
case, because $z_s(\omega_n)\sim \sqrt{|\omega_n|/\eps_T}=\pi
r\sqrt{n}/2$ is real and larger than one for finite Matsubara
frequencies, and the denominator in the expression for $i_s$
becomes large, thereby strongly suppressing the contribution of
finite $n\ne 0$.

Now we turn to the smaller rings with $r<0.6$, for which a quantum
phase transition takes place at zero temperature. We first analyze
the flux dependence of $i$ for a given temperature $T\gtrsim
T_c^0$. As long as $\varphi_{c0}$ and correspondingly $r$ do not
become very small, the same argument concerning the importance of the $\omega=0$ component that was used for rings with
$r\gtrsim 1$ in the previous paragraph is applicable here. For an estimate,
$z_s(\omega_1)\sim \sqrt{|\omega_1|/\eps_T}=\pi r/2$ becomes equal
to $1/2$ only for $r<1/\pi$, which corresponds to
$\varphi_{c0}\approx 0.27$. Interestingly, the same parameter $\pi
r/2$ determines the relevance of the nonsingular contribution in
this case (see appendix \ref{app:cplane}). As long as this parameter
does not become considerably smaller than one, it is therefore
safe to concentrate on the $\omega=0$ term of the singular
contribution only. Whenever it is justified to use only this term
near $T_c^0$, then it is also justified for larger temperatures
(as can be seen by comparing $z_s$ to $z_0$ defined in appendix~\ref{app:cplane}).

\begin{figure}
\centerline{\includegraphics[width=8cm]{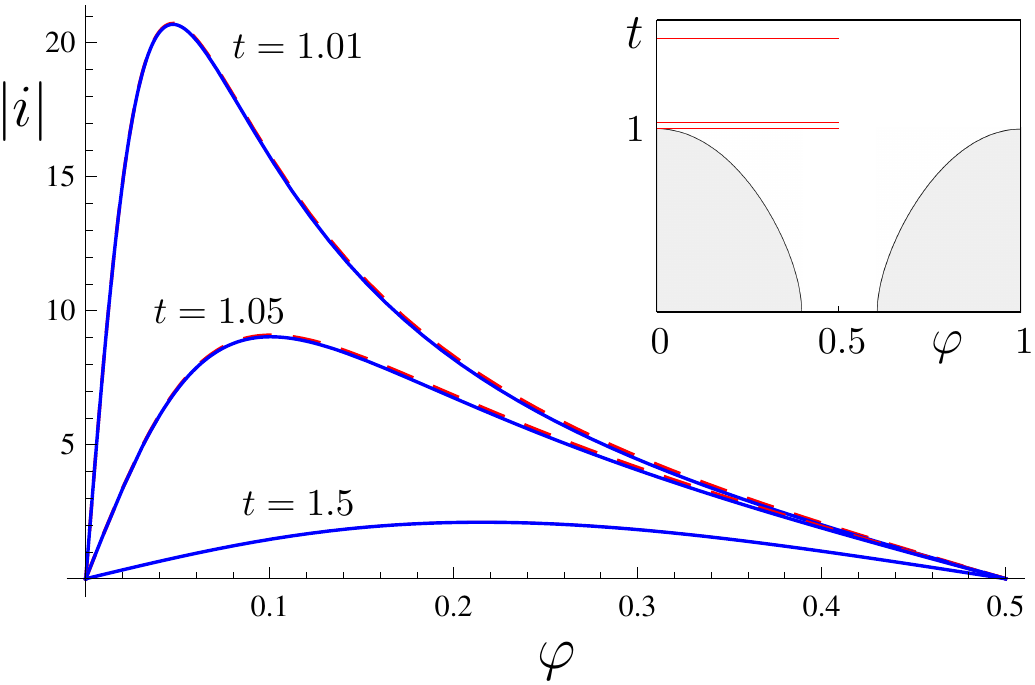}}
\caption{The modulus of the fluctuation persistent current $i=I/[T_c^0/\phi_0]$ (see Eqs.~\ref{eq:igaussgeneral} and \ref{eq:fluctprop}) is shown for a ring with $\varphi_{c0}=0.4$, corresponding to $r\approx 0.48$, for three temperatures above $T_c^0$ [$t=T/T_c^0$] (solid lines). The dashed lines show the thermal ($\omega=0$) part of $i_s$, see Eq.~(\ref{eq:thermalhighT}), which gives a very good approximation. The current decreases as the temperature grows. At the same time the shape changes considerably as explained in the text. In the inset the phase diagram is displayed, indicating the temperatures in question.}
\label{fig:HighT}
\end{figure}

To summarize this somewhat technical discussion, for not too small
rings with $\varphi_{c0}\gtrsim 0.3$ we obtain a very good description
for the entire temperature range $T>T_c^0+Gi$ by the
formula\cite{Scales} \be i\approx-2\pi t
\frac{\sin(2\pi\varphi)}{\cosh(2\pi
\sqrt{-2\alpha_c(T)/\eps_T})-\cos(2\pi \varphi)}.
\label{eq:thermalhighT} \ee The formula given above remains valid
for $T\lesssim T_c^0$ for fluxes, for which the ring is in the
normal regime. In Fig. (\ref{fig:HighT}) we display the flux
dependence of the persistent current for different temperatures.
At small fluxes, the persistent current is proportional to
$\varphi$. For temperatures close to $T_c^0$, this leads to a
rapid increase of $|i|$ for small $\varphi$. As the flux
increases, however, $T_c(\varphi)$ decreases rapidly for the small
rings under consideration here. Therefore the distance to the
critical line in the phase diagram grows as $\varphi$ increases
while the temperature is kept constant. This is why the current
subsequently drops. For larger temperatures, the situation
is different. As the temperature increases, $\cosh(2\pi
z_s)\approx \exp(2\pi z_s)$ grows and the flux dependence is
determined by the numerator in the expression for $i_s$,
Eq.~(\ref{eq:ismain}). As a consequence the shape becomes more
sinusoidal. For an estimate we can use that for $T\gtrsim T_c^0$,
$z_s\sim \sqrt{\eps}r$, see Eq.~(\ref{eq:Ambgauss}). The
exponential decay of $i$ and transition to a sinusoidal shape
therefore starts at $\eps\approx 1/(2\pi r)^2$. The persistent
current at these high temperatures results from a combined effect
of many angular momentum modes and is therefore much less
sensitive to the rapid drop of the transition line than for
temperatures $T\sim T_c^0$.

\subsection{Fluctuation persistent current near the quantum critical point and at intermediate temperatures for rings with $r<0.6$}

The situation is quite different in the low temperature limit
$T\ll T_c^0$ for rings with $r<0.6$, to which we will turn now.
Indeed, for very low temperatures a restriction to the thermal
fluctuations, namely those with $\omega=0$, is not justified as we
will see now.

We discuss the vicinity of the critical line for $T\ll T_c^0$.
Here one can expand the denominator in the general expression \ref{eq:ismain} for
$i_s$ in small
\be
\Delta\varphi_T=[\varphi-\varphi_c(T)]/\varphi_c(T)
\ee as well as
small $|\omega|/\alpha_c(T)$. In this way one obtains the
approximate relation\cite{Scales} \be
i_s\sim-\frac{2\varphi_c(T)}{\gamma_E\varphi_{c0}^2}h(\Delta\varphi_T,t),
\label{eq:isgauss} \ee where the dimensionless function $h$ is
defined as \be \quad h(\Delta\varphi_T,t)=\pi
T\sum_\omega\frac{1}{|\omega|+4\alpha_c(T)\Delta\varphi_T}.
\label{eq:hsum} \ee The same expression can be obtained directly
from the initial formula for $i$ ($i=I/(T_c^0/\phi_0)$ with $I$ given in
Eq.~\ref{eq:igaussgeneral}), if one identifies $\varphi\approx
\varphi_c(T)$, and considers the most singular angular
momentum mode $n=0$ only. It is worth noting, however, that for
fixed $n$ the sum in $\omega$ is ultraviolet divergent (even
\emph{before} expanding in $|\omega|/\alpha_{c}(T)$). When
proceeding in this way a cut-off has therefore to be introduced by
hand. In contrast, our formula (Eq.~\ref{eq:ismain}) for $i_s$ immediately reveals that
terms in the sum with $\omega>2\alpha_c(T)$ are suppressed, since
$z_s$ becomes real and $\cosh(2\pi z_s)$ grows rapidly for larger
Matsubara frequencies. We can therefore perform the sum with
logarithmic accuracy and choose $\omega=2\alpha_c(T)$ as the upper
cut-off. Put in different words, the upper limit for the
$|\omega|$-summation is effectively provided by mutual
cancelations between different angular momentum modes.

The result of the described procedure is
\be
\label{eq:hexact}
h(\Delta\varphi_T,t)=\frac{1}{2s}+\psi\left(1+\frac{s}{2\Delta\varphi_T}\right)-\psi\big(1+s\big),
\ee
where
\be
s=\frac{2\Delta\varphi_T}{t}\frac{\alpha_c(T)}{\pi T_c^0}\approx \frac{\Delta\varphi_T}{\gamma_E t}.
\ee
The first term in the expression for $h$ is the classical $\omega=0$ contribution to the sum. These thermal fluctuations are proportional to the temperature and correspondingly vanish for $T\rightarrow 0$. This does, however, not imply the vanishing of the persistent current in this limit. In order to see this more clearly, let us display the asymptotic behavior of the function $h$:
\be
h&\sim&\left\{\ba{cc}\frac{1}{2s}+\ln\frac{1}{2\gamma_E
t}&\Delta\varphi_T\ll t\ll 1\\\ln\frac{1}{\Delta\varphi_T}&t\ll
\Delta\varphi_T\ll1\ea\right. .
\ee
Even at vanishing temperatures, a flux dependent contribution to the persistent current $i_s\propto \ln(1/\Delta\varphi)$ remains. Close to the critical mean field line (see Fig.~\ref{fig:Tc}) there is a parametrically large enhancement of the persistent current due to quantum
fluctuations that decays slowly when moving away from that line.

\begin{figure}
\centerline{\includegraphics[width=8cm,clip=]{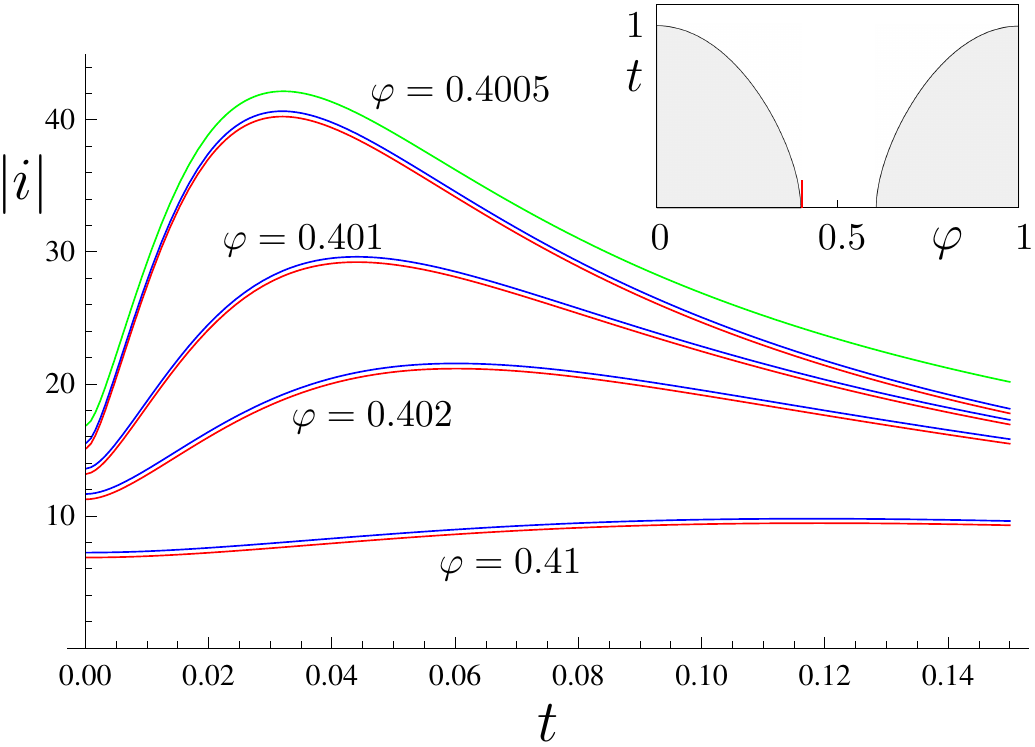}}
\caption{The modulus of the persistent current $i=I/[T_c^0/\phi_0]$ (see Eqs.~\ref{eq:igaussgeneral} and \ref{eq:fluctprop}) in the vicinity of the quantum critical point is shown in red as a function of the reduced temperature $t=T/T_c^0$ for four different fluxes $\varphi$ close to $\varphi_{c0}=0.4$. The inset shows the mean field phase diagram indicating the region relevant for this plot. The persistent current decreases when moving away from the quantum critical point. It displays a pronounced maximum at low but finite temperatures. Thermal fluctuations grow with increasing temperature, but for fixed flux the system moves further away from the critical line. At vanishing temperatures the persistent current is still large and entirely caused by quantum fluctuations. Also shown is $i_s$ (of Eq.~(\ref{eq:ismain})) in blue, which provides a very good approximation. The small difference between $i$ and $i_s$ is $i_{ns}$ (Eq.~\ref{eq:insgeneral}). In this temperature regime $i_{ns}$ can be obtained from Eq.~(\ref{eq:inslowT}). The green line was calculated from the approximate result of Eq.~(\ref{eq:isgauss}), which provides a handy estimate for the magnitude of $i$ and the position of the maximum near the quantum critical point. }
\label{fig:QCP}
\end{figure}

In order to put this result into perspective, it is instructive to compare to the persistent current in normal metal rings. The magnitude of the normal persistent current is\cite{Scales,Scurrent}
\be
I_{N}\sim
\frac{1}{\phi_0}\frac{D}{R^2}\frac{1}{\log{g}}.
\ee
The asymptotic behavior of $h$ for $t\ll
\Delta\varphi_T\ll1$ implies, that the persistent current due to pair fluctuations near $\varphi_{c0}$ is parametrically larger and at low $T\ll T_{c}^0$ given by
\be
I_{\rm FL}\approx - \frac{T^0_c}{\phi_0}\;\frac{1}{\varphi_{c0}}\;\frac{\xi}{R}\log\left(\frac{1}{\Delta\varphi}\right)\label{eq:ifl},
\ee
where $\Delta\varphi\equiv (\varphi-\varphi_{c0})/\varphi_{c0}$ measures the distance to the critical flux $\varphi_{c0}$. Since $r^{-1}=\xi/R$ is a number of order 1 and $\frac{D}{R^2}=\frac{8}{\pi}\frac{T_c^0}{r^2}$ for a weakly disordered superconductor,
we find an enhancement factor of $\log(g)\log(1/\Delta\varphi)$. When increasing the flux at fixed temperature the persistent current decays logarithmically away from the transition. It should be kept in mind, however, that the persistent current vanishes at $\varphi=0.5$ due to symmetry reasons (see the discussion below Eq.~\ref{eq:igaussgeneral}). The validity of the approximations leading to Eq. \ref{eq:hexact} is restricted to small $\Delta \varphi_T$, for larger $\Delta\varphi_T$ the full expression for $i_s$ should be used (see Fig.~\ref{fig:QCP}).

Turning to finite temperatures next, it is important that $\Delta\varphi_T$ is $T$-dependent itself and therefore, in order to reveal the full
$T$-dependence of $i_s$, one should first find the transition line $\alpha_c(T)$. We display the persistent current near the quantum critical point in Fig.~\ref{fig:QCP}, also comparing the different approximations and showing the contribution of the classical zero frequency part in the sum of Eq.~(\ref{eq:hsum}).

The maximum of $|i|$ at finite $T$ is a result of two competing mechanisms. As $T$ grows from zero, thermal fluctuations become stronger. At the same time the distance to the critical line becomes larger for fixed fixed $\varphi$, which eventually leads to a decrease of $|i|$. With the help of the following analytic approximation,
\be
\varphi^2_c(T)\sim\varphi^2_{c0}(1-2\gamma_E^2t^2/3) \quad (T\ll T_c^0),
\ee
one can obtain an estimate for the position of the maximum $\varphi_m(T)$ in the $\varphi-T$ phase diagram,
\be
\label{eq:peaktpose}
\frac{\varphi_{m}(T)-\varphi_{c0}}{\varphi_{c0}}=\left(1-\sqrt{1-{16\gamma_Et}/{3}}\right)\frac{\gamma_Et}{4}-\frac{1}{3}\gamma_E^2t^2.\no\\
\ee

\begin{figure}
\centerline{\includegraphics[width=8cm]{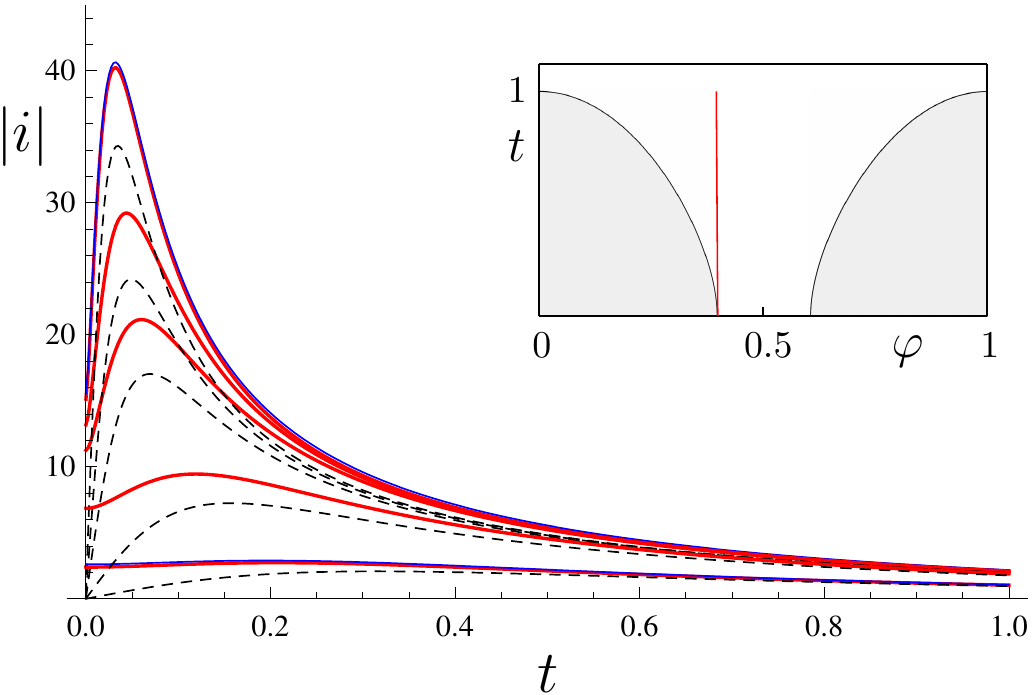}}
\caption{The modulus of the persistent current $i=I/[T_c^0/\phi_0]$ (calculated numerically according to Eq.~\ref{eq:igaussgeneral} in combination with Eq.~\ref{eq:fluctprop}) is shown in red as a function of the reduced temperature $t=T/T_c^0$ for three different fluxes $\varphi$ close to $\varphi_{c0}=0.4$, $\varphi=0.4005$, $0.401$, $0.402$ $0.41$ and $0.45$ from top to bottom. The inset shows the mean field phase diagram indicating the region relevant for this plot. Also displayed is the thermal ($\omega=0$) contribution to the current, which goes to zero at vanishing temperatures (black dashed line). Thermal fluctuations become increasingly important as the temperature grows. For two examples we indicate the singular contribution $i_s$ (of Eq.~(\ref{eq:ismain})) in blue, which gives a very good approximation for all parameters. The small difference between $i$ and $i_s$ is $i_{ns}$.}
\label{fig:QC}
\end{figure}

In appendix~\ref{app:cplane} it is shown that at low temperatures $T\ll T_c^0$ the nonsingular contribution to the persistent current $i_{ns}$ can be written as
\be
i_{ns}=2\pi t\sum_\omega \int_{-\infty}^\infty dx\; \frac{H(\varphi_{cT}\sqrt{\mbox{e}^x+{|\omega|}/{2\alpha_{cT}}},\varphi)}{x^2+\pi^2}\quad
\label{eq:inslowT}
\ee
where $H(x,\varphi)=\sin(2\pi\varphi)/(\cosh(2\pi x)-\cos(2\pi \varphi))$.
One can perform the integration in $\omega$ at zero temperature
\be
i_{ns}&=&\frac{\eps_T}{T_c^0}\int_0^\infty dy\;\;H(\sqrt{y},\varphi)\no\\
&&\times\Big(1+({2}/{\pi})\arctan\left[{\ln( y/\varphi^2_{c0})}/{\pi}\right]\Big)
\label{eq:inszeroT}
\ee
It is obvious, that $i_s$ and $i_{ns}$ have opposite signs. For a comparison with the singular contribution near the critical point it is instructive to calculate the zero temperature value of $i_{ns}$ right at the critical flux, $i_{ns}(\varphi_{c0}=0.45)=0.17$ and $i_{ns}(\varphi_{c0}=0.4)=0.43$. Since $i_{ns}$ is also a monotonously decreasing function of $\varphi$ and vanishes at $\varphi=0.5$ we conclude that it is numerically small for all fluxes of our interest, $|i_{ns}|\ll |i_s|$. The same remains true at finite temperatures, see Figs.~\ref{fig:QCP}, \ref{fig:QC}.

We display the temperature dependence of the persistent current in
the entire temperature interval $0<T<T_c^0$ in Fig.~\ref{fig:QC}.
For comparison, the thermal $\omega=0$ contribution is also shown.
One can see, that at low temperatures nonzero Matsubara
frequencies give a sizable contribution to the persistent current.
When increasing the temperature the thermal $\omega=0$
contribution becomes increasingly important. In Fig.~\ref{fig:QC}
we compare the numerically obtained current $i$ to the
approximation $i_s$. Obviously, it provides a very good
approximation in the entire temperature range.

In summary, in this section we analyzed the persistent current near the flux-tuned quantum critical point and then extended the discussion to the temperature regime $0<T<T_c^0$. Even at vanishingly small temperatures quantum fluctuations lead to a persistent current that is parametrically larger than the normal persistent current. When increasing the flux starting from $\varphi_c$ the persistent current decreases slowly and vanishes for $\varphi=0.5$. At this point the contributions from all angular momentum modes precisely cancel. When increasing the temperature from zero at fixed flux the persistent current increases initially since thermal fluctuations set in. This can be particularly well seen from Fig.~\ref{fig:QC} (dashed line). On the other hand, the distance to the critical line increases for a fixed flux when increasing the temperature, which reduces fluctuations. The competition of these two mechanisms leads to a maximum in the persistent current at finite temperature. This maximum is the more pronounced the closer the flux is to the critical flux $\varphi_c$. We derived an approximate expression for the position of the maximum in the $\varphi-T$ phase diagram, Eq.~(\ref{eq:peaktpose}). It can be expected that the general features, such as the saturation at low $T$ and the appearance of a maximum at finite $T$ remain intact for non-ideal rings, e.g., if the rings have a finite width, because they are mainly determined by a single dominant angular momentum mode. In contrast, the critical flux $\varphi_c$ may vary (see next section) and the sign change of the current may occur at a flux that is not equal to a half-integer (since many modes are involved and the phase diagram is not perfectly periodic in $\varphi$ any more).

\section{Discussion}
\label{sec:discussion}

Our results are obtained for the case when the flux acts as a
pair-breaking mechanism. Other pair-breaking mechanisms, e.g.
magnetic impurities or a magnetic field penetrating the ring
itself will lead to similar results. They cause a reduction of
$T_c$ to zero, the pair fluctuations, however, lead to a
parametric enhancement of the persistent current in the normal state.
Ref.~\onlinecite{Bary08} suggests that a similar mechanism due to
magnetic impurities is related to the unexpectedly large persistent current in
noble metal rings.\cite{Levylong90, Ambegaokar90a}

As mentioned previously, the maximal
reduction of $T_c$ at finite flux in the experiment of
Ref.~\onlinecite{Koshnicklong07} was about 6 \%. It has
so far not been possible to measure the persistent current close
to the quantum phase transition. For this type of experiment one
would need \emph{both} sufficiently small rings [in order to
fulfill the condition $r<0.6$] \emph{and} a measurement device
that allows measuring the persistent current at the comparatively
strong magnetic fields necessary to generate a flux of
$\phi\approx \phi_0/2$ threading the small area $\pi R^2$. (In the experiment at Yale~\cite{Bleszynskijayich09}
the use of cantilever required a strong magnetic field that most probably reduces almost totally the
contributions of pair fluctuations.)

 We
suggest two possible strategies to relax these conditions. If the
size of the ring is the main problem, one can try to use wider
rings, because in this case the magnetic field penetrating the
annulus of the ring helps to suppress superconductivity. A first
consequence is that the critical flux $\varphi_{c0}$ is reduced and
correspondingly the condition $\varphi_{c0}<0.5$ for observing the
quantum phase transition in the ring geometry is less restrictive,
the ring radius $R$ is allowed to be larger. This effect, however,
is rather small, as we will see below. A second effect is that the
phase diagram is no longer periodic. It can then be advantageous
to consider the transition at $\phi\approx 1.5 \phi_0$ or even
higher fluxes, because then the effect of the magnetic field
itself (as opposed to the flux) is stronger (see
Fig.~\ref{fig:width} below). This approach requires measurements
at high magnetic fields. If in turn the main problem lies in
measuring at high magnetic fields, then the addition of magnetic
impurities can help. Magnetic impurities reduce $T_c^0$ itself and
thereby also reduce $\varphi_{c0}$. We will now discuss the two
mentioned effects in more detail.

{\it Rings of finite width:} So far we considered the idealized case for which the width $w$ of
the ring (in radial direction) is vanishingly small. Next we
discuss corrections to this result, resulting from a finite width.
While doing so, we will still assume that the width is much
smaller than both the coherence length $\xi$ and the penetration
depth $\lambda$. The first assumption implies that the order
parameter field does not vary appreciably as a function of the
radius $r$, the second assumption implies that the magnetic field
is almost constant as a function of $r$.

For a ring of finite width the pair-breaking parameter acquires a
correction, $\alpha_{0n}=\alpha_{0n}^{(0)}+\alpha_{0n}^{(1)}$.
Here, $\alpha^{(0)}_{0n}=\frac{D}{2R^2}(n-\varphi)^2$ is the
width-independent part used so far and the leading corrections in
$w/R$ are \cite{Groff68} \be
\alpha_{0n}^{(1)}=\frac{D}{2R^2}\frac{w^2}{4R^2}\left(\varphi^2+\left[\frac{1}{3}+\frac{w^2}{20R^2}\right]n^2\right)
\label{eq:alphan01} \ee The relation $\xi^2=\pi D/(8T^0_c)$ can be
used to express the result through $T_c^0$ and $r$. Most
importantly, the width-dependent correction $\alpha_{0n}^{(1)}$ is
not a function of $n-\phi/\phi_0$. Correspondingly, the phase
diagram is no longer periodic in $\varphi$. The
interpretation of this result is simple. Since superconductivity
is already weakened by the magnetic field, $T_c$ can be suppressed at a smaller flux compared to
a ring of vanishing width. For the sake of brevity, we will write
only the leading correction in the following.

Let us examine some of the consequences. For the transition line
we should now solve an equation analogous to
Eq.~(\ref{eq:tcvarphi}), where now $\alpha_0(\varphi)$ should
be replaced by $\alpha_{0,m}=\min_{n}(\alpha_{0n})$. Let us first
consider the regime of small suppression $r\gg 1$ and small fluxes
$\varphi \ll 1$. Then we can use $T_c(\varphi)\sim T_c^0$,
$\alpha_{0,m}=\alpha_{0,0}$ and for $\alpha_{0,0}(\varphi)\ll
T_c(\varphi)$ one can approximately calculate the reduction of the
transition temperature. At vanishing flux, there is no magnetic
field and $T_c(\varphi=0)=T^0_c$ is unchanged, at small but finite
flux, however, there is a correction, \be T_c(\varphi)\approx
T^0_c\left(1-\frac{\varphi^2}{r^2}\left(1+\frac{w^2}{4R^2}\right)\right)
\ee For small fluxes the $T_c$ reduction is slightly stronger than
for vanishing width.

The condition for a suppression of $T^0_c$ to zero close to
$\varphi\sim 1/2$ can also be found. As for the case $w=0$ the
critical value for $\alpha$, for which $T_c$ vanishes, is
$\alpha_{c0}=\pi T_c^0/2\gamma_E$. The critical flux, however,
should now be found by equating $\alpha_{c0}$ to
$\alpha_{0,0}(\varphi)=\frac{4}{\pi}\frac{\varphi^2}{r^2}T_c^0\left(1+\frac{w^2}{4R^2}\right)$.
The result for the critical flux is \be \varphi_{c0}
\approx 0.83\; r \left(1-\frac{w^2}{8R^2}\right) \ee As expected,
for a ring of finite width the critical flux is reduced.

In Fig.~\ref{fig:width} we show the mean field transition line for
two rings with the same radius $r=0.66$, but different widths. The
ring with vanishingly small width has a flux-periodic phase
diagram and does not show a quantum phase transition, since
$r>0.6$. The other ring has a width of $w=R/3$. One observes three
main changes. The maxima in $T_c$ at finite flux are reduced
compared to $T_c^0$. They are shifted towards smaller flux, i.e.
they do no longer occur at integer values of $\varphi$. Finally,
the ring exhibits a quantum phase transition close to
$\varphi=1.5$.
\begin{figure}
\centerline{\includegraphics[width=7cm]{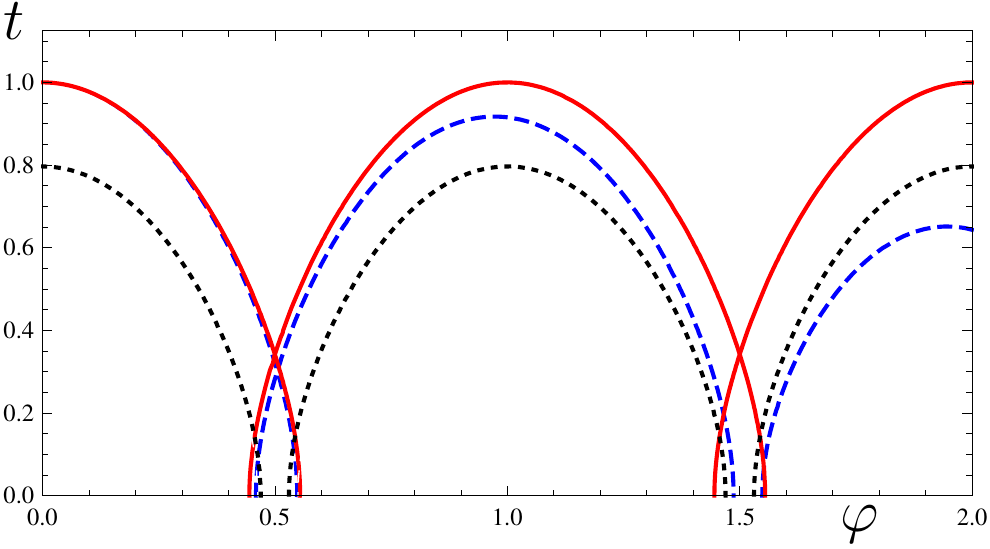}}
\caption{The mean field transition line for a ring with finite
width $w=R/3$ and $r=2/3$ (dashed blue line), for a ring with vanishingly
small width $w=0$, but same radius $r=2/3$ (solid red line), and for a ring with vanishing width $w=0$ and same radius $r=2/3$ in the presence of magnetic impurities [$1/T_c^0\tau_s=0.25$] (black dotted line). It can be seen that for the ring with finite width a quantum phase transition occurs near $\varphi=1.5$, while for the ideal one-dimensional ring there is no quantum critical point due to the periodicity of the phase diagram. In the presence of magnetic impurities the phase diagram for the ring with $w=0$ remains periodic, but now quantum transitions can be found close to $\varphi=n+1/2$ for any integer $n$.} \label{fig:width}
\end{figure}

{\it Magnetic impurities:} In this article we have so far discussed the role of an external magnetic
field as the origin of the pair-breaking mechanisms. In principle, there are other
effects that may cause pair-breaking. Among them are proximity
effect, exchange field, magnetic impurities or interaction
with the electromagnetic environment. Each pair-breaking mechanism
will have its own pair-breaking parameter, and to a good approximation the total pair breaking parameter $\alpha_{0;tot}$ is the sum of the individual ones. The
effects of these pair-breaking mechanisms can be obtained formally
by substituting $\alpha_0$ by
$\alpha_{0;tot}$ in the formulas discussed above. In particular, transition temperature $T_c(\varphi)$ can be obtained from Eq.~\ref{eq:tcvarphi} after the replacement $\alpha_0(\varphi)\rightarrow\alpha_{0,tot}(\varphi)$. The condition for the quantum critical point reads $\alpha_{0,tot}(\varphi)=\pi T_c^0/2\gamma_E$.

\begin{figure}
\centerline{\includegraphics[width=7cm]{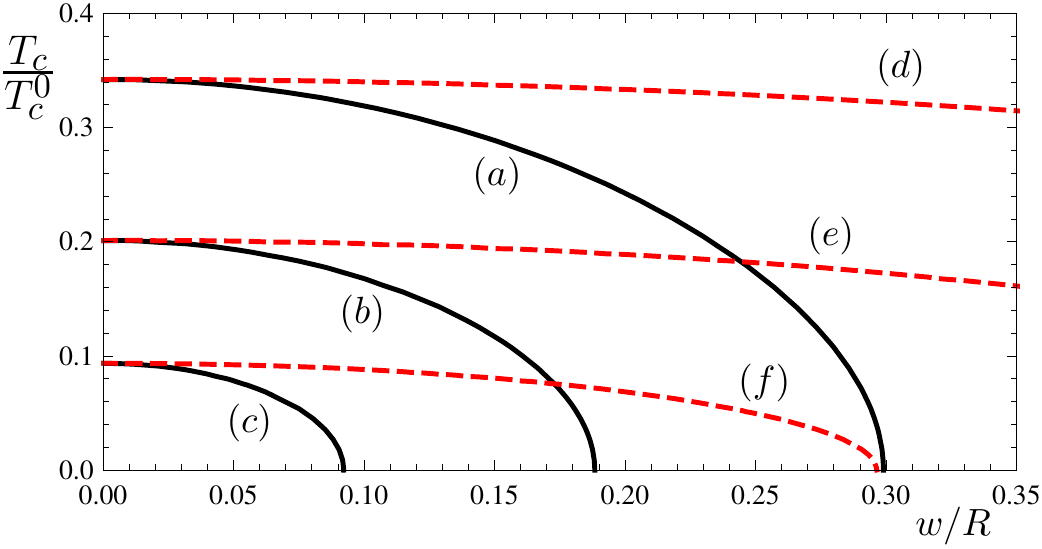}}
\caption{
The mean field transition temperature of rings with $r=2/3$ is plotted as a function of the width $w/R$ for two fluxes, $\varphi=1.5$ (black solid lines), and $\varphi=0.5$ (red dashed lines), and for a varying spin scattering rate $1/T_c^0\tau_s$ [$1/T_c^0\tau_s=0$ for $(a)$, $(d)$, $1/T_c^0\tau_s=0.1$ for $(b)$, $(e)$, and $1/T_c^0\tau_s=0.15$ for $(c)$, $(f)$]. Compared to the case $\varphi=0.5$, the width dependence is much stronger for $\varphi=1.5$. The addition of a small amount of magnetic impurities reduces the minimal width, for which $T_c$ vanishes. It is noteworthy that for a larger scattering rate than considered in this figure, a quantum transition can be induced near $\varphi=0.5$ even in the limit of vanishing width (see Fig.~\ref{fig:width})}
\label{fig:tcvswidth}
\end{figure}

\begin{figure}
\centerline{\includegraphics[width=8.5cm]{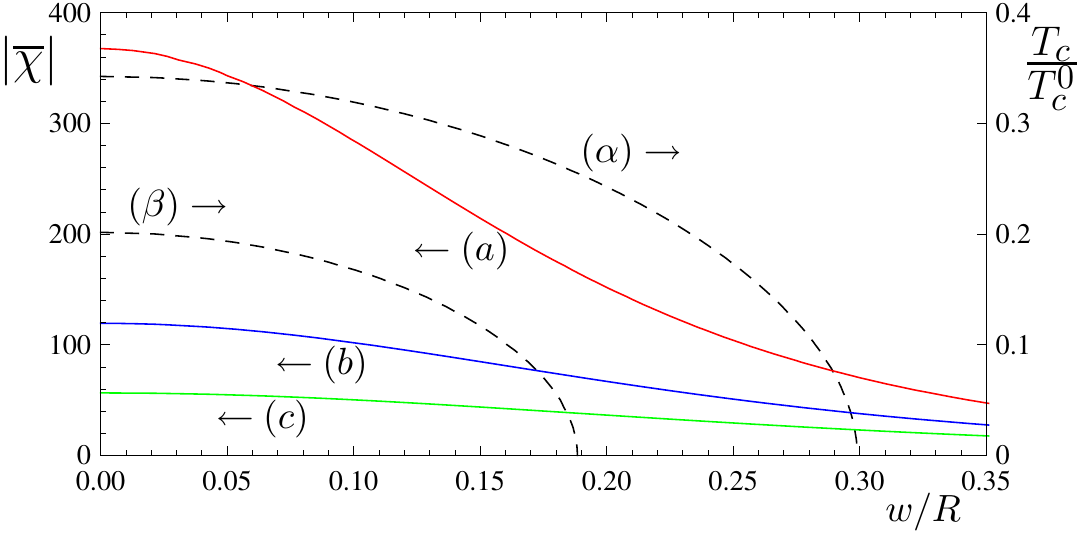}}
\caption{The modulus of the susceptibility for $\varphi=1.5$ and $T=0.5\; T_c^0$ is plotted as a function of the width $w/R$ for three values of the spin scattering rate, $1/T_c^0\tau_s=0.0$ for $(a)$, $0.1$ for $(b)$, and $0.2$ for $(c)$. The radius of the ring is $r=2/3$.
Also shown is the width-dependence of the mean field transition temperature $T_c/T_c^0$ at $\varphi=1.5$ (dashed lines) for two values of the spin scattering rate, $1/T_c^0\tau_s=0$ for $(\alpha)$, and $1/T_c^0\tau_s=0.1$ for $(\beta)$. For $1/T_c^0\tau_s=0.2$, $T_c$ is equal to zero for arbitrary width.
The susceptibility at $\varphi=1.5$ is entirely caused by fluctuations, because the mean field transition temperature is smaller than $0.5 T_c^0$, even in the limit of vanishing width. We see that the susceptibility is a smooth function of the width. In particular, it remains large even after passing the threshold values of $w/R$ for which $T_c$ vanishes.}
\label{fig:chivswidth}
\end{figure}

Of particular interest is the case of magnetic impurities,
discussed by Bary-Soroker, Entin-Wohlman and
Imry,~\cite{Bary08,Bary09} for which $\alpha_{0,mi}=1/\tau_s$ is equal to the scattering rate caused by the magnetic impurities and independent of the flux. A sufficiently large concentration of magnetic impurities
will reduce $T_c$ to zero [even at vanishing flux] and a large persistent current is
obtained due to the pair fluctuations. Ref.~\onlinecite{Bary08} suggest
that the large persistent current observed in copper rings~\cite{Levylong90} may be attributed to such pairing fluctuations.

As emphasize earlier, in our case the addition of magnetic impurities may push the system to the quantum critical point at a smaller external magnetic field, since it provides an additional flux-independent pair-breaking mechanism. If the main experimental difficulty is to perform sensitive measurements at high magnetic fields, introducing magnetic impurities might therefore make possible the experimental observation of the flux tuned quantum critical point, see Fig.~\ref{fig:tcvswidth}

\section*{Acknowledgments}
We thank E.~Altman, H. Bary-Soroker, A. I. Buzdin, O. Entin-Wohlman, H.~J.~Fink, A. M. Finkel'stein, Y.~Imry, Y.~Liu, F. von Oppen for
useful discussions, and K. Moler, N. Koshnick and H. Bluhm for
stimulating discussions and for sharing their numerical results
with us. We acknowledge financial support from the Minerva
Foundation, DIP and ISF grants.

\vspace{.75cm}
\begin{appendix}
\section{}

\label{app:twozero}

In this appendix we will show how to calculate the partition function when two modes $\psi_0$ and $\psi_1$ are taken into consideration.
We start from Eq.~(\ref{eq:2}) and introduce the notation $A_n=a_n/T_c^0$, $B=b/(2VT_c^0)$, $x=|\psi_0|^2$ and $y=|\psi_1|^2$. Then the expression for the partition function becomes
\be
\mathcal{Z}=\pi^2\int_0^\infty dxdy\;\mbox{e}^{-A_0
x-A_1y-B(x^2+y^2+4xy)}
\ee
We want to perform one integration explicitly. To this end we first change integration variables to $w=x+y$ and $z=x-y$ and obtain $\mathcal{Z}=\frac{\pi^2}{2}\int_0^\infty dw\int_{-w}^{w}dz\;\exp(-wA_+-zA_--\frac{3}{2}Bw^2+\frac{1}{2}Bz^2)$, where $A_{\pm}=(A_0\pm A_1)/2$. After changing the order of integration we find
\be
\mathcal{Z}&=&\frac{\pi^2\sqrt{\pi}}{\sqrt{3}B}\;\mbox{e}^{A_+^2/6B}\int_0^\infty dz\;\\
&&\quad\times\mbox{e}^{z^2}\cosh(A_-z\sqrt{2/B})\;\erfc(\sqrt{3}z+A_+/\sqrt{6B})\no
\ee
Next we derive a formula for persistent current. Combining the formulas $I=(2mR^2\phi_0)^{-1}\sum_{n=0,1}(n-\varphi)\left\langle|\psi_n|^2\right\rangle$ and $\left\langle|\psi_n|^2\right\rangle=-\partial_{A_n}\ln \mathcal{Z}$ one finds
\be
i=\frac{1}{4mR^2T_c^0}\left(\partial_{A_-}+2\Delta\varphi\;\partial_{A_+}\right)\ln\mathcal{Z},
\label{eq:ivialogZ}
\ee
where $\Delta\varphi=\varphi-1/2=4mR^2T_c^0A_-$. Let us note that
\be
\label{eq:derivatives}
\partial_{A_+}\ln\mathcal{Z}&=&\frac{A_+}{3B}-\frac{\pi^2\sqrt{\pi}}{12B^{3/2}\mathcal{Z}}\left[\mbox{e}^{x^2_0}\mbox{erfc}(x_0)+\mbox{e}^{x_1^2}\mbox{erfc}(x_1)\right]\no\\
\partial_{A_-}\ln\mathcal{Z}&=&-\frac{A_-}{B}-\frac{\pi^2\sqrt{\pi}}{4B^{3/2}\mathcal{Z}}\left[\mbox{e}^{x^2_0}\mbox{erfc}(x_0)-\mbox{e}^{x_1^2}\mbox{erfc}(x_1)\right]\no\\
\ee
where $x_n=A_n/(2\sqrt{B})=\eps_n/Gi$ coincides with the variable $x_n$ used in the main text. After combining these results one finds
\be
i_2=-4\frac{r^2}{Gi}\mathcal{M}_-+\frac{8}{3}x_-\mathcal{M}_+,\label{eq:appi2}
\ee
where
\be
\mathcal{M}_{\pm}=x_{\pm}\mp\frac{1}{\mathcal{P}}\left(\textrm{e}^{x_0^2}\textrm{erfc}(x_0)\pm \mbox{e}^{x_1^2}\;\erfc(x_1)\right)
\ee
and
\be
\mathcal{P}=4\int_0^\infty dz\;\textrm{e}^{3z^2+2(2x_1-x_0)z+x_1^2}\;\textrm{erfc}(2z+x_1)
\ee
where $x_{\pm}=(x_0\pm x_1)/2$. An analogous formula has been given in Ref. \onlinecite{Daumens98}.

Most interesting for us is the quantity $\overline{\chi}(1/2)=-\left.\partial i_2/\partial\varphi\right|_{\varphi=1/2}$. It is worth noting that for $\varphi=1/2$ one finds $A_-=x_-=0$ (i.e $x_0=x_1$), and formulas simplify considerably. We can obtain the formula for $\overline{\chi}(1/2)$ directly by differentiating the result for $i_2$ or by first differentiating Eq.~(\ref{eq:ivialogZ}) and then using the relations in Eq.~(\ref{eq:derivatives}). In the latter case one obtains as an intermediate step the relation $\overline{\chi}(1/2)=-\frac{\phi_0^2}{2mR^2T_c^0\mathcal{Z}}(\partial_{A_+}+\frac{1}{8T_c^0mR^2}\partial^2_{A_-})\left.\mathcal{Z}\right|_{A_-=0}$. The result for $\overline{\chi}$ is stated in the main text, Eq.~(\ref{eq:f2}).

\section{}

\label{app:cplane}
\begin{figure}
\centerline{\includegraphics[width=6.5cm]{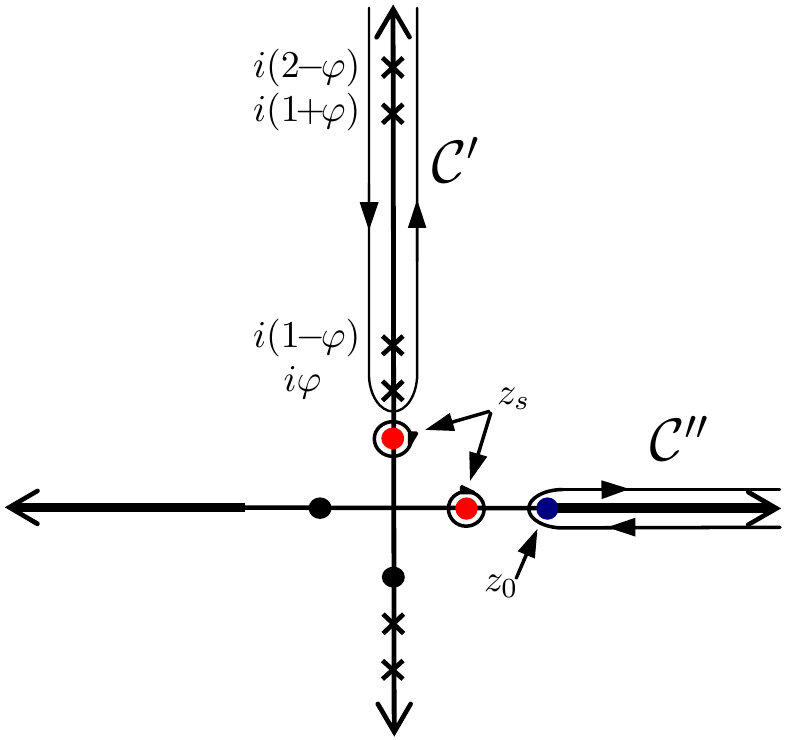}}
\caption{The complex $z$ plane. The poles of $H(z,\varphi)$ (Eq.~\ref{eq:RandH}) as a function of $z$ are indicated by crosses together with the original integration contour $\mathcal{C}'$. Equivalently, the integral in Eq.~\ref{eq:iGintegralinz} can be calculated by integrating along the path $\mathcal{C}''$, and additionally taking into account the residue at the pole $z_s$. $z_s$ can be either imaginary or real and is responsible for the dominant contribution $i_s$ to the persistent current. The contour $\mathcal{C}''$ encloses $z_n$ and $\tilde{z}_n$, the poles of $R(z,\omega)$, which are positioned on the real axis to the right of $z_0$. As the temperature decreases the distance between adjacent $z_n$ and $\tilde{z}_n$ becomes small and the sum over the corresponding residues can be approximated by an integral along a branch cut.}
\label{fig:cplane}
\end{figure}

In this appendix we sketch the derivation of the expressions for the persistent current in the Gaussian approximation used in the main text. We also make contact with Ref.~\onlinecite{Bary09}, where fluctuations at temperatures $T>T_c(\varphi=0)$ have been examined [In the presence of magnetic impurities, $T_c(\varphi=0)$ differs from $T_{c}^0$, the transition temperature in the absence of any pair-breaking mechanism, and may even vanish].
Our starting point is Eq.~(\ref{eq:igaussgeneral}),
\be
i_G=-t\sum_{n,\omega}\frac{\partial_\varphi\mathcal{L}^{-1}(n,\omega)}{\mathcal{L}^{-1}(n,\omega)}
\ee
Using standard methods for transforming the sum in $n$ into an integral in the complex plane, we can write $i$ as
\be
i_G&=&-it\sum_\omega\;\oint_{\mathcal{C}'}dz\;H(z,\varphi) \;R(z,\omega)
\label{eq:iGintegralinz}
\ee
where
\be
R(z,\omega)&=&\frac{\partial_z \tilde{\mathcal{L}}^{-1}(-iz,\omega)}{\tilde{\mathcal{L}}^{-1}(-iz,\omega)},\no\\
\quad H(z,\varphi)&=&\frac{\sin(2\pi\varphi)}{\cosh(2\pi z)-\cos(2\pi \varphi)}
\label{eq:RandH}
\ee
and we use the notation
\be
&&(\nu\tilde{\mathcal{L}})^{-1}(-iz,\omega)=\\
&&\ln\left(\frac{T}{T^0_c}\right)+\psi\left(\frac{1}{2}+\frac{|\omega|-\eps_Tz^2}{4\pi T}\right)-\psi\left(\frac{1}{2}\right).\no
\ee
The contour $\mathcal{C}'$ includes all the poles of $H(z,\varphi)$ with $\Im{z}>0$. These poles are located at
\be
z=i\varphi, i(1-\varphi), i(1+\varphi),i(2-\varphi),\dots.
\ee
The contour $\mathcal{C'}$ does not include poles of $R(z,\omega)$, see Fig.~\ref{fig:cplane}. Our goal will be to deform the contour of integration in such a way that the integral can be evaluated at the poles of $R$ with the help of the residue theorem. Poles of $R(z,\omega)$ occur either due to zeros or poles of $\tilde{\mathcal{L}}^{-1}(-iz,\omega)$. Due to the dependence of $\mathcal{L}^{-1}$ on $z^2$ the poles of $R(z,\omega)$ come in pairs, for each pole at a point $z$ there is a pole at $-z$.

Zeros of $\tilde{\mathcal{L}}^{-1}(-iz,\omega)$ can be classified according to the value of
\be
A(z,\omega)=1/2+(|\omega|-\eps_Tz^2)/(4\pi T),
\ee
the argument of the digamma function. There is a pair of zeros $\pm z_s$, for which $A$ is positive, and there is a pair of zeros $\pm z_n$ for each interval $-n-1<A<-n$, where $n=0,1,2,\dots$. The existence of such zeros is clear from the reflection formula
\be
\psi(1-z)=\psi(z)+\pi \cot(\pi z).
\ee
Poles of $\tilde{\mathcal{L}}^{-1}(-iz,\omega)$ originate from poles of the digamma function for $A=-n$, $n=0,1,2,\dots$ and we label the pairs accordingly, $\pm\tilde{z}_n$, where
\be
\tilde{z}_n=\sqrt{((2n+1)2\pi T+|\omega|)/{\eps_T}}.
\ee
Note that all $z_n$ and $\tilde{z}_n$ are real and $z_n<\tilde{z}_n<z_{n+1}$.

The zeros at $\pm z_s$ deserve special attention. For $T>T_c^0$, $z_s$ is real. For $T<T_c^0$, however, $z_s$ can be either real or imaginary. Indeed, due to the relation $\ln({T}/{T^0_c})=\psi({1}/{2})-\psi({1}/{2}+{\alpha_c(T)}/{2\pi T})$, we can write $z_s=\varphi_c(T)\sqrt{|\omega|/2\alpha(T)-1}$ for $T<T_c^0$. $z_s$ is real for $\omega>2\alpha(T)$ and purely imaginary $z_s=i\varphi_c(T)\sqrt{1-|\omega|/2\alpha(T)}$ for $\omega<2\alpha(T)$.

By deforming the contour $\mathcal{C}'$ and using the symmetry and convergence-properties of the integrand we can write $i_s$ as an integral along a contour $\mathcal{C}''$ that encloses all poles of $R(z,\omega)$ on the positive imaginary axis for $0<\Im{z}<\varphi$, and on the positive real axis in the complex $z$-plane, see Fig.~\ref{fig:cplane}. The integral can then formally be calculated with the help of the residue theorem.

When $\omega$ is small and $z_s$ imaginary, $H$ can become large for $\varphi\sim\varphi_s$, reflecting the closeness to the phase transition.
Evaluating the residue at $z_s$ one finds the singular contribution to the persistent current
\be
i_s=-2\pi t\sum_\omega H(z_s,\varphi)\label{eq:isgeneral}
\ee
The nonsingular contribution arises from all other poles of $R$, and can be written as
\be
i_{ns}=-2\pi t\sum_n\sum_{\omega} [H(z_n,\varphi)-H(\tilde{z}_n,\varphi)].\label{eq:insgeneral}
\ee

 In general, the poles $z_n$ need to be determined numerically. The derived representation is particularly useful in two limiting cases, either for very high or very low temperatures. For very high temperatures $z_s$ is real the spacing between consecutive $z_s, z_n, \tilde{z}_n$ is large. Since $H(z,\varphi)$ decays fast as a function of $z$ one can approximate the result well by considering just the smallest of the $z_s, z_n, \tilde{z}_n$ and in this way obtain relatively simple formulas. This approximation was utilized in Ref.~\onlinecite{Bary09}, where a similar representation was derived for temperatures $T>T_c(\varphi=0)$ (in the presence of magnetic impurities), when all poles of $R(z,\omega)$ are real. Another useful limit is the limit of very low temperatures, when $z_s$ is possibly imaginary, but the other $z_n$, $\tilde{z}_n$ are closely placed on the real axis. Then, calculating the residues for $i_{ns}$ is very similar to an integration along a branch cut as we will describe now. In this low temperature limit one can write
\be
(\nu\mathcal{L})^{-1}_{n,\omega}&=&\psi\left[\frac{1}{2}+\frac{\alpha_n+\frac{|\omega|}{2}}{2\pi T}\right]-\psi\left[\frac{1}{2}+\frac{\alpha_c(T)}{2\pi T}\right]\no\\
&\stackrel{T\ll T_c^0}{\approx}&\ln\left[\frac{\alpha_n+\frac{|\omega|}{2}}{\alpha_c(T)}\right]
\ee
Following the same analysis presented above with this approximate representation, $i_s$ remains unchanged. For $i_{ns}$ we have to perform an integration along a branch cut appearing for $z>z_0\sim\sqrt{|\omega|/\eps_T}$, which is related to the branch cut of the logarithm. It gives the result
\be
i_{ns}=2\pi t\sum_\omega \int_{0}^\infty \frac{dy}{y}\;\frac{H(\varphi_c(T)\sqrt{y+|\omega|/2\alpha_c(T)},\varphi)}{\ln^2y+\pi^2}\quad
\ee
Simple substitution gives the formula Eq.~(\ref{eq:inslowT}) stated in the main text. In the limit $T\rightarrow 0$, it is more convenient to use the relation $[y(\ln^2y+\pi^2)]^{-1}=\frac{1}{\pi}\partial_y\arctan\left(\ln(y)/\pi\right)$, perform a partial integration in $y$, subsequently use $\partial_yf(y+|\omega|/2\alpha_c(T))=2\alpha_c(T)\partial_{|\omega|}f(y+|\omega|/2\alpha_c(T))$ to perform the integral in $\omega$ and combine the result with the boundary term. The result is Eq.~(\ref{eq:inszeroT}).

Finally, let us make contact with the analysis of Ref.~\onlinecite{Bary09}. In this paper, fluctuations were analyzed for high temperatures $T>T_c(\varphi=0)$ (in the presence of magnetic impurities) and the flux harmonics $i_m$ of the persistent current $i=\sum_m i_m\;\sin(2\pi m\varphi)$ were calculated. In this situation one may use the Poisson summation formula to perform the sum over angular momentum modes, since the flux dependence of $i$ is smooth for $T\gg T_c(\varphi=0)$. [This is not so for $T<T_c(\varphi=0)$ due to the phase transition to the superconducting state, or, in a mathematical language, due to the presence of the pole at purely imaginary $z_s$.] A comparable situation arises for $T>T_c^0$ in the absence of magnetic impurities: All poles of $R(z,\omega)$ (including $z_s$) are real, one can use Eqs.~\ref{eq:isgeneral} and \ref{eq:insgeneral} and the relation
\be
H(x,\varphi)=2\sum_{m=1}^\infty\;\mbox{e}^{-2\pi x m} \sin(2\pi m\varphi)
\ee
which is valid for $x>0$, to write
\be
i&=&-4\pi t\sum_{m=1}^{\infty}\;\sin(2\pi m\varphi)\;\\
&&\times\sum_{\omega}\;\left[\mbox{e}^{-2\pi z_s m}+\sum_n\left(\mbox{e}^{-2\pi z_n m}-\mbox{e}^{-2\pi \tilde{z}_n m}\right)\right],\no
\ee
from which one can read off the flux harmonics $i_m$.

\end{appendix}


\begin{thebibliography}{38}
\bibitem[{Sca()}]{Scales}
\bibinfo{note}{Some important scales involved in the problem and notation we
  use: The zero temperature coherence length for disordered superconductors is
  $\xi^2=\pi D/8T_c^0$, where $T_c^0$ is the transition temperature at
  vanishing flux. We measure the radius of the ring $R$ in units of $\xi$:
  $r=R/\xi$. The reduced radius $r$ is related to the critical flux
  $\varphi_{c0}$ (at which the quantum phase transition occurs for sufficiently
  small rings) by $r^2=8\gamma_E\varphi^2_{c0}/\pi^2$, where $\gamma_E\approx
  1.78$ [$\ln\gamma_E=-\psi(1/2)-\ln 4$]. The Thouless energy $\eps_T=D/R^2$
  and $T_c^0$ are related by $\eps_T/T_c^0=\pi/\gamma_E\varphi_{c0}^2=8/{\pi
  r^2}$. We often measure the persistent current $I$ in units of
  $T_c^0/\phi_0$, where $\phi_0=\pi/e$ is the superconducting flux
  quantum\cite{Units}, $I=iT_c^0/\phi_0$, and in a similar way write for the
  susceptibility $\chi=\overline{\chi}T_c^0/\phi_0^2$. The zero-dimensional
  Ginzburg parameter $Gi$ is given as $Gi=\sqrt{7\zeta(3)/4\pi^2\nu T_c^0 V}$,
  where $\nu$ is the density of states at the Fermi level and $V$ is the volume
  of the ring. The parameter $\Lambda=1/r^2Gi$ determines whether a description
  in terms of a few modes only is a good approximation in the critical regime
  (for $\Lambda \gg 1$) or not. Defining the dimensionless conductance of the
  ring as $g=R_Q/R_\circ=2e^2\nu D R_Q V/(2\pi R)^2$, $R_Q=\pi /e^2$, one can
  find an alternative expression, $\Lambda\approx 5\sqrt{g}/r$.}
\end{thebibliography}
\end{document}